\documentclass[12pt]{article}
\usepackage{fancyhdr}

\usepackage{mathrsfs}
\usepackage[T1]{fontenc}
\usepackage{mathpazo}
\usepackage{setspace}
\usepackage{amsfonts}
\usepackage{amssymb}
\usepackage{amsmath}
\usepackage{epsfig}
\usepackage{latexsym}
\usepackage{color}
\usepackage{graphicx}
\usepackage{nicefrac}
\usepackage[latin1]{inputenc}
\usepackage{slashed}
\usepackage{multirow}
\usepackage{comment}
\usepackage{soul}
\usepackage{hyperref}
\usepackage{cite}

\usepackage[titletoc]{appendix}

\def\hybrid{\topmargin -20pt    \oddsidemargin 0pt
        \headheight 0pt \headsep 0pt
        \textwidth 6.25in       
        \textheight 9.25in       
        \marginparwidth .875in
        \parskip 5pt plus 1pt   \jot = 1.5ex}

\hybrid

\def\baselinestretch{1.2}

\catcode`\@=11

\def\marginnote#1{}
\newcount\hour
\newcount\minute
\newtoks\amorpm
\hour=\time\divide\hour by60
\minute=\time{\multiply\hour by60 \global\advance\minute by-\hour}
\edef\standardtime{{\ifnum\hour<12 \global\amorpm={am}%
        \else\global\amorpm={pm}\advance\hour by-12 \fi
        \ifnum\hour=0 \hour=12 \fi
        \number\hour:\ifnum\minute<10 0\fi\number\minute\the\amorpm}}
\edef\militarytime{\number\hour:\ifnum\minute<10 0\fi\number\minute}

\def\draftlabel#1{{\@bsphack\if@filesw {\let\thepage\relax
   \xdef\@gtempa{\write\@auxout{\string
      \newlabel{#1}{{\@currentlabel}{\thepage}}}}}\@gtempa
   \if@nobreak \ifvmode\nobreak\fi\fi\fi\@esphack}
        \gdef\@eqnlabel{#1}}
\def\@eqnlabel{}
\def\@vacuum{}
\def\draftmarginnote#1{\marginpar{\raggedright\scriptsize\tt#1}}

\def\draft{\oddsidemargin -.5truein
        \def\@oddfoot{\sl preliminary draft \hfil
        \rm\thepage\hfil\sl\today\quad\militarytime}
        \let\@evenfoot\@oddfoot \overfullrule 3pt
        \let\label=\draftlabel
        \let\marginnote=\draftmarginnote
   \def\@eqnnum{(\theequation)\rlap{\kern\marginparsep\tt\@eqnlabel}%
\global\let\@eqnlabel\@vacuum}  }

\def\preprint{\twocolumn\sloppy\flushbottom\parindent 2em
        \leftmargini 2em\leftmarginv .5em\leftmarginvi .5em
        \oddsidemargin -.5in    \evensidemargin -.5in
        \columnsep .4in \footheight 0pt
        \textwidth 10.in        \topmargin  -.4in
        \headheight 12pt \topskip .4in
        \textheight 6.9in \footskip 0pt
        \def\@oddhead{\thepage\hfil\addtocounter{page}{1}\thepage}
        \let\@evenhead\@oddhead \def\@oddfoot{} \def\@evenfoot{} }

\def\numberbysection{\@addtoreset{equation}{section}
        \def\theequation{\thesection.\arabic{equation}}}

\def\underline#1{\relax\ifmmode\@@underline#1\else
        $\@@underline{\hbox{#1}}$\relax\fi}

\def\titlepage{\@restonecolfalse\if@twocolumn\@restonecoltrue\onecolumn
     \else \newpage \fi \thispagestyle{empty}\c@page\z@
        \def\thefootnote{\fnsymbol{footnote}} }

\def\endtitlepage{\if@restonecol\twocolumn \else \newpage \fi
        \def\thefootnote{\arabic{footnote}}
        \setcounter{footnote}{0}}  

\catcode`@=12
\relax

\def\figcap{\section*{Figure Captions\markboth
        {FIGURECAPTIONS}{FIGURECAPTIONS}}\list
        {Figure \arabic{enumi}:\hfill}{\settowidth\labelwidth{Figure
999:}
        \leftmargin\labelwidth
        \advance\leftmargin\labelsep\usecounter{enumi}}}
 \relax
\def\tablecap{\section*{Table Captions\markboth
        {TABLECAPTIONS}{TABLECAPTIONS}}\list
        {Table \arabic{enumi}:\hfill}{\settowidth\labelwidth{Table
999:}
        \leftmargin\labelwidth
        \advance\leftmargin\labelsep\usecounter{enumi}}}
 \relax
\def\reflist{\section*{References\markboth
        {REFLIST}{REFLIST}}\list
        {[\arabic{enumi}]\hfill}{\settowidth\labelwidth{[999]}
        \leftmargin\labelwidth
        \advance\leftmargin\labelsep\usecounter{enumi}}}
 \relax
\makeatletter
\newcounter{pubctr}
\def\publist{\@ifnextchar[{\@publist}{\@@publist}}
\def\@publist[#1]{\list
        {[\arabic{pubctr}]\hfill}{\settowidth\labelwidth{[999]}
        \leftmargin\labelwidth
        \advance\leftmargin\labelsep
        \@nmbrlisttrue\def\@listctr{pubctr}
        \setcounter{pubctr}{#1}\addtocounter{pubctr}{-1}}}
\def\@@publist{\list
        {[\arabic{pubctr}]\hfill}{\settowidth\labelwidth{[999]}
        \leftmargin\labelwidth
        \advance\leftmargin\labelsep
        \@nmbrlisttrue\def\@listctr{pubctr}}}
 \relax
\makeatother
\newskip\humongous \humongous=0pt plus 1000pt minus 1000pt

\newif\ifdtup

\relax

\def\be{\begin{equation}}
\def\ee{\end{equation}}
\def\ba{\begin{eqnarray}}
\def\ea{\end{eqnarray}}

\def\del{\partial}

\def\k{\kappa}

\def\b{\beta}

\def\g{\gamma}

\def\d{\delta}
\def\D{\Delta}

\def\m{\mu}

\def\l{\lambda}

\def\s{\sigma}

\def\no{\noindent}

\def\qq{\qquad}

\def\IR{\relax{\rm I\kern-.18em R}}

\def \ha {{1\over 2}}

\def \ov {\over}

\def\IR{\relax{\rm I\kern-.18em R}}
\def\IL{\relax{\rm I\kern-.18em L}}

\def\inv{^{\raise.15ex\hbox{${\scriptscriptstyle -}$}\kern-.05em 1}}

\def\Tr{{\rm Tr}}

\begin{document}

\renewcommand{\theequation}{\thesection.\arabic{equation}}
\csname @addtoreset\endcsname{equation}{section}

\newcommand{\beq}{\begin{equation}}
\newcommand{\eeq}[1]{\label{#1}\end{equation}}
\newcommand{\ber}{\begin{equation}}
\newcommand{\eer}[1]{\label{#1}\end{equation}}
\newcommand{\eqn}[1]{(\ref{#1})}
\begin{titlepage}
\begin{center}


${}$
\vskip .2 in

{\large\bf  All-loop correlators of integrable $\lambda$-deformed $\sigma$-models}

\vskip 0.4in

{\bf George Georgiou,$^1$ Konstantinos Sfetsos}$^{2}$\ and\ {\bf Konstantinos Siampos}$^{3}$
\vskip 0.1in

\vskip 0.1in
{\em
${}^1$Institute of Nuclear and Particle Physics,\\ National Center for Scientific Research Demokritos,\\
Ag. Paraskevi, GR-15310 Athens, Greece
}
\vskip 0.1in

 {\em
${}^2$Department of Nuclear and Particle Physics,\\
Faculty of Physics, National and Kapodistrian University of Athens,\\
Athens 15784, Greece\\
}
\vskip 0.1in

{\em ${}^3${Albert Einstein Center for Fundamental Physics,\\
Institute for Theoretical Physics / Laboratory for High-Energy Physics,\\
University of Bern,
Sidlerstrasse 5, CH3012 Bern, Switzerland
}}
\vskip 0.1in

{\footnotesize \texttt georgiou@inp.demokritos.gr, ksfetsos@phys.uoa.gr, siampos@itp.unibe.ch}


\vskip .5in
\end{center}

\centerline{\bf Abstract}

\no 
We compute the 2- and 3-point functions of currents and primary
fields of $\lambda$-deformed integrable $\sigma$-models characterized also
by an integer $k$. Our results apply for any semisimple group $G$,
for all values of the deformation parameter $\lambda$ and up to
order $1/k$. We deduce the OPEs and equal-time commutators of all
currents and primaries. We derive the currents' Poisson brackets
which assume Rajeev's deformation of the canonical structure of the
isotropic PCM, the underlying structure of the integrable
$\lambda$-deformed $\sigma$-models.
We also present analogous results in two limiting cases of special interest,
namely for the non-Abelian T-dual of the PCM and for the pseudodual model.

\vskip .4in
\noindent
\end{titlepage}
\vfill
\eject

\newpage

\tableofcontents

\noindent

\def\baselinestretch{1.2}
\baselineskip 20 pt
\noindent


\setcounter{equation}{0}
\section{Introduction and motivation}
\renewcommand{\theequation}{\thesection.\arabic{equation}}

One of the most intriguing conjectures in modern theoretical physics is the AdS/CFT correspondence \cite{Maldacena:1997re} which,
in its initial form, states the equivalence between type-IIB superstring theory on the $AdS_5 \times S^5$ background and the
maximally supersymmetric field theory in four dimensions, i.e. $\mathcal{N}=4$ SYM.
In recent years, a huge progress has been made in calculating physical observables employing both sides of the duality.
These calculations
managed to probe the strongly coupled regime of the gauge theory which is practically unaccessible by other means.
The key feature that
allowed this progress is integrability. $\mathcal{N}=4$ SYM from one side and the two-dimensional
$\s$-model from the other, are believed to be integrable order by order in perturbation theory. It is clear that one
way to construct generalizations of the original AdS/CFT scenario is to try to maintain the key property of integrability.

\no
The aim of this work is to study the structure of a class of two-dimensional $\s$-models,
the so-called $\lambda$-deformed models constructed in \cite{Sfetsos:2013wia}.
For isotropic couplings the deformation is integrable in the group case and in the symmetric and semi-symmetric coset cases
\cite{Sfetsos:2013wia,Itsios:2014vfa,Hollowood:2014rla,Hollowood:2014qma} (for the $su(2)$ group case integrability is
preserved for anisotropic, albeit diagonal couplings \cite{Sfetsos:2014lla}).
They are also closely related \cite{Klimcik:2002zj,Klimcik:2008eq,Vicedo:2015pna,Hoare:2015gda,Sfetsos:2015nya,Klimcik:2015gba}
to the so-called $\eta$-deformed models for group and coset spaces introduced
in \cite{Klimcik:2002zj,Klimcik:2008eq} and in
\cite{Delduc:2013fga,Delduc:2013qra,Arutyunov:2013ega}, respectively.
This relation is via Poisson--Lie T-duality
and an analytic continuation of coordinates and of the parameters of the $\s$-models \cite{Hoare:2015gda,Sfetsos:2015nya,Klimcik:2015gba}.
There are also embedings of the $\l$-deformed models as solutions of supergravity \cite{Sfetsos:2014cea,Demulder:2015lva,Borsato:2016zcf}.

\no
In particular, we shed light into the structure of the $\l$-deformed models
by computing the two- and three-point functions of all currents and operators
exactly in the deformation parameter and up to order $1/k$. This work is based and further extends symmetry ideas and techniques
originated in our previous work in \cite{Georgiou:2015nka}. The results of this work are summarized in section \ref{conclusion}.

Our starting point is the WZW action
\be
S_{{\rm WZW},k}(g) = -{k\ov 4\pi} \int\text{d}^2\sigma\, {\rm Tr}(g^{-1} \del_+ g g^{-1} \del_- g)
+ { k\ov 24\pi} \int_B {\rm Tr}(g^{-1}\text{d}g)^3\ ,
\label{WZW}
\end{equation}
for a generic semisimple group $G$, with $g\in G$ parametrized by $X^\m$, $\m=1,2,\dots ,\dim G $.
We will use the representation matrices $t_a$ which obey
the commutation relations $[t_a,t_b]=f_{abc} t_c$ and are normalized as ${\rm Tr}(t_a t_b)=\delta_{ab}$.
These matrices are taken to be Hermitian and therefore the Lie-algebra structure constants $f_{abc}$ are purely
imaginary.
The chiral and anti-chiral currents are defined as
\be
J^a_+ = -i\, {\rm Tr}(t_a \del_+ g g^{-1}) = R^a_\m \del_+ X^\m \ ,\qq J^a_- = -i\,
{\rm Tr}(t_a g^{-1} \del_- g )= L^a_\m \del_- X^\m\ .
\label{jjd}
\end{equation}
The left and right invariant forms $L^a= L^a_\m \text{d}X^\m $ and $R^a= R^a_\m \text{d}X^\m $  are related as
\be
R^a = D_{ab}L^b \ ,\qq D_{ab}={\rm Tr}(t_a g t_b g^{-1})\ .
\label{jjdd}
\end{equation}

\no
We are interested in the non-Abelian Thirring model action (for a general
discussion, see \cite{Dashen:1974gu,Karabali:1988sz}), namely the WZW two-dimensional
conformal field theory (CFT) perturbed by a set of classically marginal operators which are bilinear in the currents
\be
\label{WZW-pert}
S=S_{{\rm WZW},k}(g) +{k\ov 2\pi} \sum_{a,b=1}^{\dim G} \lambda_{ab} \int \text{d}^2\s\,  J_+^a J_-^b\ ,
\end{equation}
where the couplings are denoted by the constants $\lambda_{ab}$.
An action having the same global symmetries as \eqn{WZW-pert},
and to which reduces for small values of $\lambda_{ab}$ has been derived in
\cite{Sfetsos:2013wia} (see also \cite{Balog:1993es} for the $SU(2)$ case), by gauging a common symmetry subgroup of an action involving the PCM model and the WZW actions.
It reads \cite{Sfetsos:2013wia}
\begin{equation}
S_{k,\l}(g) = S_{{\rm WZW},k}(g) +{k\ov 2\pi} \int \text{d}^2\s\, J_+^a (\l^{-1} - D^T)^{-1}_{ab} J_-^b\ ,
\label{efff}
\end{equation}
where we have assembled in a general real matrix $\lambda$ the coupling constants $\lambda_{ab}$.
In addition, this action, as well as \eqn{WZW-pert}, is invariant under the generalized parity transformation
\begin{equation}
\label{parity}
\sigma^\pm\mapsto\sigma^\mp\ ,\qquad g\mapsto g^{-1}\ ,\qquad \lambda\mapsto\lambda^T\ .
\end{equation}

\no
The $\beta$-functions for the running of couplings under
the Renormanization Group (RG) flow using \eqn{efff} were computed in \cite{Itsios:2014lca,Sfetsos:2014jfa}
and completely agree with the computation of the same RG-flow equations using CFT techniques based on \eqn{WZW-pert}
in  \cite{Kutasov:1989dt} for a single (isotropic) coupling, i.e. when $\lambda_{ab}=\lambda \delta_{ab}$
and in \cite{Gerganov:2000mt} for symmetric $\lambda_{ab}$.
Based on that it was conjectured in \cite{Itsios:2014lca,Sfetsos:2014jfa} that \eqn{efff} is the effective action for \eqn{WZW-pert}
 valid to all orders in $\lambda$ and up to order $1/k$.
In the same works it was realized that \eqn{efff} has the remarkable symmetry
\begin{equation}
S_{-k,\l^{-1}}(g^{-1}) = S_{k,\l}(g)\ .
\label{dualsym}
\end{equation}
This has been instrumental in computing the anomalous dimensions of
currents for the isotropic case exactly in $\lambda$ and up to order in $1/k$ \cite{Georgiou:2015nka}
and will be central in the present work as well.
We should stress that this is not a symmetry of the non-Abelian Thirring model action \eqn{WZW-pert}.
However, using path integral techniques and special properties of the WZW model action, it was argued in \cite{Kutasov:1989aw}
that the effective action of the non-Abelian Thirring model (not known at the time) should be invariant
under the above duality-type symmetry $(\l,k)\mapsto (\l^{-1},-k)$ (for $k\gg 1$).

\section{The set up}

\subsection{OPE's at the conformal point}

In what follows, we shall need the operator product expansion (OPE) of the currents in the
Euclidean regime with complex coordinates $z=\ha(\tau + i \s)$
and $\bar z$. For the holomorphic ones the singular part of their OPE reads \cite{Witten:1983ar,Knizhnik:1984nr}
\begin{equation}
\label{OPE}
J^{a}(z)J^{b}(w)=\frac{f_{abc}}{\sqrt{k}} {J^c(w)\ov  z-w} + \frac{\delta_{ab}}{(z-w)^2}
\end{equation}
and similarly for the OPE between the antiholomorphic currents $\bar J^a(\bar z)$.
Of course the OPE $J^{a}(z)\bar{J}^b(w)$ is regular.
The difference from the more conventional form of these OPE's arises because
we have rescaled the currents as $J^a\mapsto J^a/\sqrt{k}$ which suits our purposes
since in that way, as will shall see, we keep easily track of the contributions of various terms to the correlators of the perturbed theory.

\no
The CFT contains affine primary fields $\Phi_{i,i'}(z,\bar z)$ transforming in the
irreducible representations $R$ and $R'$, with matrices $t_a$ and $\tilde t_a$,
under the action of the currents $J^a$ and $\bar J^a$,
so that $i=1,2,\dots , \dim R$ and $i'=1,2,\dots , \dim R'$.
Specifically,
\begin{equation}
\label{jjj}
\begin{split}
& J_a(z) \Phi_{i,i'}(w,\bar w) = -{1\ov \sqrt{k}} {(t_a)_i{}^j \Phi_{j,i'}(w,\bar w)\ov z-w}\ ,
\\
&\bar J_a(z) \Phi_{i,i'}(w,\bar w) = {1\ov \sqrt{k}} {(\tilde t_a)^{j'}{}_{i'} \Phi_{i,j'}(w,\bar w)\ov \bar z- \bar w}\ .
\end{split}
\end{equation}
These fields are also Virasoro primaries with holomorphic and
antiholomorphic dimensions \cite{Knizhnik:1984nr}
\begin{equation}
\D_R = {c_R\ov 2k+ c_G}\ ,\qq \bar \D_{R'} = {c_{R'}\ov 2k+ c_G}\ ,
\label{ddcft}
\end{equation}
where $c_R$, $c_{R'}$ and $c_G$ are the quadratic Casimir operators, all non-negative, in the representations $R$, $R'$ and the
adjoint representation for which $(t_a)_{bc}= f_{abc}$. They are defined as
\begin{equation}
(t_a t_a)_i{}^j = c_R \delta_i{}^j \ ,\qq (\tilde t_a \tilde t_a)_{i'}{}^{j'}
= c_{R'} \delta_{i'}{}^{j'}\ ,\qq  f_{acd} f_{bcd} = - c_G \delta_{ab} \ .
\end{equation}
In our calculations we will need the basic two- and three-point functions for these fields.
For the currents they are given by
\begin{equation}
\langle J_a(z_1) J_b(z_2)\rangle ={\delta_{ab}\ov z_{12}^2}\ ,\qq
\langle J_a(z_1) J_b(z_2) J_c(z_3)\rangle ={1\ov \sqrt{k}} {f_{abc}\ov z_{12} z_{13} z_{23}}\ ,
\end{equation}
where we employ the general notation $z_{ij}= z_i-z_j$.
We will also use the four-point function
\begin{equation}
\label{jp4}
\begin{split}
& \langle J^a(x_1) J^{a_1}(z_1) J^{a_2}(z_2) J^{a_3}(z_3)\rangle  = {1\ov k} {f_{a_1 a c} f_{c a_2 a_3}\ov
(z_1-x_1)(x_1-z_2)(x_1-z_3)(z_1-z_3)}
\\
&\hskip 5 cm + {\delta_{a a_1}\delta_{a_2 a_3}\ov (x_1-z_1)^2 (z_2-z_3)^2} +  \text{cyclic\ in}\ 1,2,3\ .
\end{split}
\end{equation}
Similar expressions hold for the antiholomorphic currents as well.
Correlators involving both holomorphic and anti-holomorphic currents vanish at the conformal point.
However, as we shall see, this will not be the case in the deformed theory.

\no
The corresponding correlators for the affine primaries are
\begin{equation}
\langle \Phi^{(1)}_{i,i'}(z_1,\bar z_1) \Phi^{(2)}_{j,j'}(z_2,\bar z_2)\rangle
={\delta_{ij}\,\delta_{i'j'}\ov z_{12}^{2\D_R} \ \bar z_{12}^{2 \bar \D_{R'}}}\ ,
\end{equation}
where the superscripts { signify} the fact that the representations for the different primaries
in correlation functions could be, in general, different.
However, for the two-point functions the two representations should
in fact be conjugate to each other for the holomorphic and anti-holomorphic sectors separately.
As such, they have the same conformal dimensions.
Recalling that the matrices $t_a$ and $\tilde t_a$ are Hermitian and
after removing the superscripts by relabeling the representation matrices we have that
\begin{equation}
{\rm Reps\ (1)\ and\  (2)\ conjugate}:\
t^{(1)}_a= t_a\ , \quad \tilde t^{(1)}_a= \tilde t_a \ ,\quad t^{(2)}_a= -t_a^*\ , \quad \tilde t^{(2)}_a= -\tilde t_a^*\ .
\label{conjj}
\end{equation}
The minus sign in the definition of the conjugate representation is very important for the matrices
to obey the same Lie-algebra.
It will turn out that, in the deformed theory, for { correlation functions} involving
two primaries to be non-vanishing, their corresponding representations must be conjugate to each other, as well.

\no
Next, consider three affine primaries transforming in the representations
$(R_i,R'_i)$, $i=1,2,3$. Then the three-point function for them is given by
\begin{equation}
\langle \Phi^{(1)}_{i,i'}(z_1,\bar z_1) \Phi^{(2)}_{j,j'}(z_2,\bar z_2)  \Phi^{(3)}_{k,k'}(z_3,\bar z_3) \rangle
={C_{ii',jj',kk'}\ov z_{12}^{\D_{12;3}}\ z_{13}^{\D_{13;2}}\ z_{23}^{\D_{23;1}}\
\bar z_{12}^{\bar \D_{12;3}}\ \bar z_{13}^{\bar \D_{13;2}}\ \bar z_{23}^{\bar \D_{23;1}}}\ ,
\end{equation}
where
\begin{equation}
\D_{12;3}=\D_{R_1}+\D_{R_2}-\D_{R_3}\ ,\qq  \bar \D_{12;3}=\bar\D_{R'_1}+\bar \D_{R'_2}-\bar \D_{R'_3}\ .
\end{equation}
and cyclic permutations of $1,2$ and $3$ for the rest.
The structure constants $C_{ii',jj',kk'}$ depend on the representations
and implicitly also on $k$. They obey various properties arising mainly from the global
group invariance of the correlation functions, which will be mentioned below in the computation of the three-point
functions involving only affine primaries.

\no
Finally, we have the three-point functions with one current and two primaries. They are given by
\begin{equation}
\langle J_a(z) \Phi^{(1)}_{i,i'}(x_1,\bar x_1) \Phi^{(2)}_{j,j'}(x_2,\bar x_2)\rangle
= -{1\ov \sqrt{k}} {(t_a\otimes \mathbb{I}_{R'})_{ij,i'j'}\ov x_{12}^{2 \D_R}\ \bar x_{12}^{2 \bar \D_{R'}}}
\left({1\ov z-x_1}-{1\ov z-x_2}\right)
\label{JJf1}
\end{equation}
and
\begin{equation}
\langle \bar J_a(\bar z) \Phi^{(1)}_{i,i'}(x_1,\bar x_1) \Phi^{(2)}_{j,j'}(x_2,\bar x_2)\rangle
= {1\ov \sqrt{k}} { (\mathbb{I}_{R}\otimes \tilde t_a^*)_{ij,i'j'}\ov x_{12}^{2 \D_R}\ \bar x_{12}^{2 \bar \D_{R'}}}
\left({1\ov \bar z-\bar x_1}-{1\ov \bar z- \bar x_2}\right)\ ,
\label{JJf2}
\end{equation}
where we have used the fact that, for a non-vanishing result,
the representations in which the primaries transform have to be conjugate
to each other for the holomorphic and the antiholomorphic sectors, separately.
Also $\mathbb{I}_{R}$ and $\mathbb{I}_{R'}$ are the identity elements for the corresponding representations.

\no
Correlators with two currents and one affine primary field are zero at the conformal point and  will remain zero in the deformed
theory as well.

\subsection{Symmetry and correlation functions}

In order to compute the correlation functions of currents and of primary fields
we will heavily use the symmetry of the effective action for the non-Abelian Thirring model \eqn{dualsym}.
First let's consider correlation
functions for currents only. At the conformal point when $\l=0$ the currents are given in terms of the group element by \eqn{jjd}
and are, of course, chirally and anti-chirally conserved on shell.
Obviously, in the deformed theory these currents will be dressed and
will receive $\lambda$-corrections. One expects that since their definition contains derivatives
there will be operator ambiguities at the quantum
level.
We propose that these dressed currents are given by
\begin{equation}
\begin{split}
& J^a_+(g)_{k,\l} = -{i\ov 1+\l} (\mathbb{I}-\l D)^{-1}_{ab} \Tr(t^b \del_+ g g^{-1})\ , \quad
\\
&J^a_-(g)_{k,\l} = {i\ov 1+\l} (\mathbb{I}-\l D^T)^{-1}_{ab} \Tr(t^b g^{-1}\del_- g )\ .
\end{split}
\label{jredf}
\end{equation}
These become the correct chiral and anti-chiral currents when $\l=0$ (up to a minus sign for
$J_-$).
Also, they are components of an on shell conserved current.
The attentive reader will notice that the dressed current components in \eqn{jredf} are nothing, but,
up to a factor of $\lambda$,
the gauge fields evaluated on-shell in the original construction
of  \eqn{efff} in \cite{Sfetsos:2013wia} by a gauging procedure.
Hence, it is natural to consider correlation functions of the $J_\pm^a$'s as defined above.
In addition, we have that
\begin{equation}
J_\pm^a(g^{-1})_{-k,\l^{-1}} = \l^2 J^a_\pm(g)_{k,\l} \ .
\label{jjgh}
\end{equation}

\no
Passing to the Euclidean regime
we have for the two-point function of the holomorphic component of the currents that
\begin{equation}
\langle J^a(x_1) J^b(x_2)\rangle_{k,\l} ={1\ov Z_{k,\l}} \int {\cal D}[g] J^a(g(x_1))_{k,\l} J^b(g(x_2))_{k,\l} e^{- S_{k,\l}(g) }\ ,
\end{equation}
with the partition function being
\begin{equation}
Z_{k,\l}= \int {\cal D}[g]  e^{- S_{k,\l}(g) } = \int {\cal D}[g^{-1}]  e^{- S_{k,\l}(g^{-1}) } = 
{\int {\cal D}[g]  e^{- S_{-k,\l^{-1}}(g) }} = Z_{-k,\l^{-1}}\ .
\end{equation}
where we have used the symmetry of the action \eqn{dualsym} and the fact that the measure of integration
is invariant under $g\mapsto g^{-1}$, i.e. ${\cal D}[g^{-1}]={\cal D}[g]$.\footnote{
The measure of integration contains the Haar measure for the semisimple group $G$ which is certainly invariant under $g\mapsto g^{-1}$,
but also the factor $\det(\lambda^{-1}-D^T)$ arising from integrating out the gauge fields
in the path integral \cite{Sfetsos:2013wia}.
This can be easily seen to transform under $g\mapsto g^{-1}$ and $\lambda\mapsto \lambda^{-1}$ as (for a general matrix $\l$):
$ \det(\lambda^{-1}-D^T)\mapsto (-1)^n\det\lambda\times\det(\lambda^{-1}-D^T)$, with $n=\dim G$
and where we have used the property $D(g^{-1})=D^T(g)$.
This extra constant overall factor cancels out by the same factor arising from the partition function in the denominator in all correlation
functions.
}
Hence, the partition function of the deformed theory is invariant under the duality-type symmetry.
In addition
\be
\begin{split}
& \int {\cal D}[g] J^a(g(x_1))_{k,\l} J^b(g(x_2))_{k,\l} e^{- S_{k,\l}(g) }
\\
&
\qq = \int {\cal D}[g^{-1}] J^a(g^{-1}(x_1))_{k,\l} J^b(g^{-1}(x_2))_{k,\l} e^{- S_{k,\l}(g^{-1}) }
\\
&
\qq = {1\ov \l^4} \int {\cal D}[g] J^a(g(x_1))_{-k,\l^{-1}} J^b(g(x_2))_{-k,\l^{-1}} e^{- S_{-k,\l^{-1}}(g) } \ .
\end{split}
\ee
where we have also employed \eqn{jjgh}. Hence, we obtain that the correlation function should obey the
non-trivial identity
\begin{equation}
\l^2 \langle J^a(x_1) J^b(x_2)\rangle_{k,\l} = \l^{-2} \langle J^a(x_1) J^b(x_2)\rangle_{-k,\l^{-1}}
\end{equation}
This identity between current correlators is straightforwardly extendable to
higher order correlators involving currents with any type of currents, $J^a$'s or $\bar J^a$'s.
\begin{equation}
\l^{n+m} \langle J^{a_1} \dots J^{a_n} \bar J^{b_1} \dots \bar J^{b_m}\rangle_{k,\l} =
\l^{-n-m} \langle J^{a_1} \dots J^{a_n} \bar J^{b_1} \dots \bar J^{b_m}\rangle_{-k,\l^{-1}}\ .
\end{equation}
The overall factors of $\lambda$ can be absorbed by redefining the currents in \eqn{jredf} by a factor of $\lambda$.
In the following we assume that this is the case which implies also the absence of the factor of $\l^2$ in the r.h.s. of \eqn{jjgh}.

\no
The above conclusion for the current correlators  is in full agreement with \cite{Kutasov:1989aw}
who reached the same conclusion using the non-Abelian Thirring
model action and certain special properties of the WZW action path integral.
The advantage of employing the effective action is that one can employ the duality-type symmetry on correlation
functions involving primary fields in the deformed theory which has not been considered before.
For these fields we have that,
under the inversion of the group element the primary field $\Phi^{(1)}$ transforms to its conjugate $\Phi^{(2)}$.
Explicitly, we have that
\begin{equation}
{
\Phi^{(1)}_{i,i'}(g^{-1}) = \Phi^{(2)}_{i',i}(g)}\ ,
\label{ffsym}
\end{equation}
which means that for the representation matrices we have
\begin{equation}
t^{(1)}\leftrightarrow\tilde t^{(2)}\ ,\qq t^{(2)}\leftrightarrow\tilde t^{(1)}\ .
\end{equation}

\no
Note that if the inversion of $g$ is followed by the $\sigma\mapsto -\sigma$, i.e. the parity transformation \eqn{parity}, then
\begin{equation}
t^{(1)} \leftrightarrow  -\tilde t^{(2)}\ ,\qq t^{(2)} \leftrightarrow  -\tilde t^{(1)}\ ,
\end{equation}
and in addition the $J_a$'s and $\bar J_a$'s are interchanged.

\subsection{The non-Abelian and pseudodual chiral limits}

Besides the small $\lambda_{ab}$ limit, leading to \eqn{WZW-pert}, there are two other interesting limits of the action \eqn{efff}.
They will be instrumental in our computation of correlation functions.

\no
In the first limit \cite{Sfetsos:2013wia} one expands the matrix and group elements near the identity as
\begin{equation}
\lambda_{ab} =\delta_{ab} - {E_{ab} \ov k}  + {\cal O}\left(1\ov k^2\right)\ , \quad g = \mathbb{I} + i { v_a t^a \ov k} + {\cal O}\left( 1\ov k^2 \right)  \ ,
\label{laborio}
\end{equation}
where $E$ is a general $\text{dim}G$ square matrix. This leads to
\begin{equation}
J_\pm^a = {\del_\pm v^{a}\ov k} + {\cal O}\left( 1\ov k^2 \right)  \ ,\quad
D_{ab} = \delta_{ab}+\frac{f_{ab}}{k} + {\cal O}\left( 1\ov k^2 \right) \ ,\quad {  f_{ab} = - i f_{abc} v^c}\ .
\label{orrio}
\end{equation}
{ 
Note that our structure constants are purely imaginary so that $f_{ab}$ are indeed real.}
In this limit the action \eqn{efff} becomes
\begin{equation}
S_{\rm non\!-\!Abel}(v) = {1\ov 2 \pi} \int \text{d}^2\s\, \del_+ v^a (E +f)^{-1}_{ab}\del_- v^b  \ ,
\label{nobag}
\end{equation}
which is the the non-Abelian T-dual with respect to the $G_{L}$ action of the $\sigma$-model given by  the PCM action
with general coupling matrix $E_{ab}$.
We note that in this limit the WZW term in \eqn{efff} does not contribute at all.

\no
To discuss the second new limit, we first recall that the original derivation of the action \eqn{efff} leads for
compact groups to the restriction $0<\l<1$. However, once we have the action we may allow $\lambda$ to take values beyond this range. For instance,
the symmetry \eqn{dualsym} clearly requires that. Here in order to take a new limit we will extend the range of $\lambda$ to negative values.
We will also need the following equivalent form of the action \eqn{efff} given, after some manipulations needed to combine the
quadratic part of the WZW action and the deformation term in \eqn{efff}, by
\begin{equation}
\label{effective.action}
\begin{split}
& S_{k,\lambda}(g)=\frac{k}{4\pi}\int  \text{d}^2\s\,J^a_+\left[(\l^{-1}-D^T)^{-1}(\l^{-1}+D^T)D\right]_{ab}J_-^b
\\
& \qq \qq { -\frac{ik}{48\pi}}\int_B f_{abc}\,L^a\wedge L^b \wedge L^c\ .
\end{split}
\end{equation}
where we 
we remind the reader that our structure constants are purely imaginary. 

\no
Then we take the limit
\begin{equation}
\lambda_{ab}= -\delta_{ab} +{E_{ab} \ov k^{1/3}}\ ,\qq g= \mathbb{I} + i {v^a t^a\ov k^{1/3}}  + \dots\ ,\qq k\to \infty\ ,
\label{limps}
\end{equation}
where again $E$ is a general $\text{dim}G$ square matrix. The various quantities expand as in \eqn{orrio} with $k$ replaced by $k^{1/3}$.
Then the action \eqn{effective.action} becomes
\begin{equation}
S_{\rm pseudodual}={1\ov 8\pi} \int\text{d}^2\sigma\, \del_+ v^a  \del_- v^b\left(E_{ab} +\frac{1}{3} f_{ab}\right) \ .
\label{psac}
\end{equation}
We see that $E$ can be taken to be symmetric since any antisymmetric piece leads to a total derivative.
This action for $E_{ab} = \delta_{ab}/b^{2/3}$ is nothing by the pseudodual model action \cite{Nappi:1979ig}. Note that
the quadratic part of the WZW action and the deformation term in \eqn{efff} are equally important for
the limit \eqn{limps} to exist since each term separately diverges when this limit is taken.

\no
Since the above non-Abelian and pseudodual limits exist at the action level, we expect that
physical quantities such as the $\b$-function and the anomalous dimensions of various operators should
have a well {defined} limit as well. This will be an important ingredient in our method of computation.

\subsection{The regularization method and useful integrals}

In the Euclidean path integral the action appears as $e^{-S}$.
The action we will be using is that of the non-Abelian Thrirring model action and will be expanding around the WZW CFT part
of it. This is not in contrast with the approach of the last subsection where \eqn{efff} was used,
the reason being that the latter is the effective action of the non-Abelian Thrirring model.
Hence, it contains all $\lambda$-corrections and can be considered as a starting point to find at the
quantum level corrections in $1/k$.
Schematically, to ${\cal O}(\l^n)$, the correlation function for a number of some generic
fields $F_i$, $i=1,2,\dots $, involves the sum of expressions of the type
\be
\begin{split}
& \langle F_1(x_1,\bar x_1) F_2(x_2,\bar x_2)\dots  \rangle_\l^{(n)} = {1\ov n!} \left(-\frac{\l}{\pi}\right)^n
\int \text{d}^2z_{1\dots n} \langle J^{a_1}(z_1)\dots J^{a_n}(z_n)
\\
& \qq\qq \phantom{xxxxx} \bar J^{a_1}(\bar z_1)\dots \bar J^{a_n}(\bar z_n) F_1(x_1,\bar x_1) F_2(x_2,\bar x_2)\dots\rangle\ ,
\end{split}
\ee
where $\text{d}^2z_{1\dots n}:=\text{d}^2z_1\dots \text{d}^2z_n$ and for convenience we have dropped $k$ from our notation
in the correlation functions $\langle\cdots\rangle_{k,\l}$ of the deformed theory.

\no
That way one encounters
multiple integrals which need to be {regularized}. Our prescription to do so consists of two steps:

\no
$\bullet $ We choose the order of integration from left to right $\text{d}^2z_{1\dots n}$ and never
permute this order. This is due to the fact that due to the divergences appearing, the various
integrations are not necessarily {commuting.}

\no
$\bullet $ Internal points cannot coincide with external ones.
This means that the domain of integration is
\be
D_n = \{(z_1, z_2,\dots , z_n)\in \mathbb{C}_n :|z_i - x_j|> \varepsilon, \varepsilon>0\}\ ,\quad  \forall\ i,j\ .
\ee
However, internal points can coincide.
Also contact terms, arising from coincident external points will be allowed. The latter is a choice we
make and not a part of the regularization scheme.\footnote{All these imply that we will have
for the $\d$-functions arising in performing the various integrations that
$$
\d^{(2)}(z_i-x_j)\to 0 \ ,\quad  \d^{(2)}(z_i-z_j)\ (\text{kept})\ ,\quad \d^{(2)}(x_i-x_j)\ (\text{kept})\ ,
\quad \forall i, j\ .
$$
Note also that in the regularization of \cite{Candu:2012xc} no two points, internal or external,
can coincide and therefore all $\d$-functions arising in integrations are set to zero.
In contrast in \cite{Konechny:2010nq} all such $\d$-functions are kept.
The advantage of our regularization is that
the symmetry of the correlation functions under $k\to -k$ and $\l\to \l^{-1}$
is manifest whereas for the others it is hidden.
}
\no
We shall need the very basic integral given by
\begin{equation}
\label{id1}
\int  \frac{\text{d}^2z}{(x_1-z)(\bar{z}-\bar{x}_2)}=\pi\ln{|x_{12}|^2}\ .
\end{equation}
Clearly, if the domain of integration allows, the integral diverges for large distances.
The above result is valid provided that
the integration is performed in a domain of characteristic size $R$, e.g. a disc of radius $R$,
with the external points $x_1$ and $x_2$ excluded and in addition obeying $R\gg |x_1|, |x_2|$.
The latter conditions are responsible for the translational invariance and the reality of the result.
Even then we have to make the replacement $|x_{12}|^2\to |x_{12}|^2/R^2$ on the right hand side of \eqn{id1}.
However, in our computations there will be integrals of the same kind but with opposite sign and
$x_1$ equal to $x_2$ and which will have a small distance regulator $\varepsilon$.
Hence the factor $R$ will drop out at the end, leaving the ratio $\displaystyle \ln {\varepsilon^2\ov |x_{12}|^2}$.
This means that in practice the domain of integration is $\mathbb{R}^2$
except for the points $x_{1,2}$ which are excluded.
By appropriately taking derivatives we also have the useful integrals
\begin{equation}
\label{id2}
\int \frac{\text{d}^2z}{(x_1-z)^2(\bar{z}-\bar{x}_2)}=-\frac{\pi}{x_{12}}\ ,
\qq
\int \frac{\text{d}^2z}{( x_1- z)(\bar{z}-\bar{x}_2)^2}=-\frac{\pi}{\bar x_{12}}
\end{equation}
and
\begin{equation}\label{id3}
\int \frac{\text{d}^2z}{(x_1-z)^2(\bar{z}-\bar{x}_2)^2}=\pi^2\delta^{(2)}(x_{12})\,.
\end{equation}
In appendix \ref{vint} we have collected results for some useful to this work
integrals. We single out
\begin{equation}
\begin{split}
& \int {\text{d}^2 z\ov (z-x_1)(z-x_2)(\bar z-\bar x_1)}=  -{\pi \ov x_{12}} \ln {\varepsilon^2\ov |x_{12}|^2} \ ,
\\
&
\int {\text{d}^2 z\ov (z-x_1)(\bar z-\bar x_1)(\bar z-\bar x_2)}=  -{\pi \ov \bar x_{12}} \ln {\varepsilon^2\ov |x_{12}|^2}
\label{id22}
\end{split}
\end{equation}
and
\begin{equation}
\int\frac{\text{d}^2z}{(z-x_1)(\bar z-\bar x_2)}\,\ln |z-x_1|^2=
-\frac\pi2\ln^2 |x_1-x_2|^2\ .
\label{logpla}
\end{equation}
which are valid under the assumptions spelled out below \eqn{id1}.

\section{Current correlators}

In this section, we will focus on the two- and three-point functions involving purely currents.
These will be computed up to order $1/k$ and exactly in the deformation parameter $\lambda$.
To establish our method, employed already in \cite{Georgiou:2015nka}, as clearly as possible we first start with the computation of the two-point functions
which enables to compute the $\b$-function and the anomalous dimensions for the currents known already
from using CFT methods in \cite{Kutasov:1989dt,Gerganov:2000mt,Georgiou:2015nka} and from gravitational computations \cite{Itsios:2014lca,Sfetsos:2014jfa}.
Then we proceed to correlators involving three currents.

\subsection{Two-point functions}\label{sec:jj2}

On general grounds the correlator of $J^a$ and $J^b$  takes the form
\begin{equation}
\label{generic.corr}
\langle J^a(x_1)J^b(x_2)\rangle_\l=\d^{ab} {G_0(k,\l) \ov x_{12}^2}
\left(1 +\gamma^{(J)} \ln {\frac{\varepsilon^2}
{|x_{12}|^{2}}}\right)+\cdots\ .
\end{equation}
The result to ${\cal O}(1/k)$ and ${\cal O}(\l^3)$ was computed in sec. 2 of \cite{Georgiou:2015nka} and
reads
\begin{equation}
\langle J^a(x_1)J^b(x_2)\rangle
=\frac{\delta_{ab}}{x_{12}^2}\Big(1 - 2 {c_G\ov k}\l^3
+{ c_G\ov k} (\lambda^2-2 \lambda^3)
\ln{\frac{\varepsilon^2}{|x_{12}|^2}} +{ \frac{1}{k}\, {\cal O}(\lambda^4)}\Big)\ .
\label{jjll}
\end{equation}
Comparing with the general form of the two-point function \eqn{generic.corr}
we have that
\begin{equation}
G_0(k,\l)  = 1 - 2 {c_G\ov k}\left(\l^3 + {\cal O}(\lambda^4)\right)
\end{equation}
and
\begin{equation}
\g^{(J)} = {c_G\ov k}\left(\l^2 - 2  \l^3  + {\cal O}(\lambda^4)\right)  \ .
\label{gammapert1}
\end{equation}

\no
Similarly the correlator of $J^a$ and $\bar J^b$ should assume the form
\begin{equation}
\label{generic.corr2}
\begin{split}
& \langle J^a(x_1)\bar J^b(x_2)\rangle_\l=\d^{ab} {\tilde G_0(k,\l) \ov |x_{12}|^2} \left( 1+\gamma^{(J)} \ln {\frac{\varepsilon^2}
{|x_{12}|^{2}}}\right)
\\
 &\qq \qq\qq + \delta_{ab} \d^{(2)}(x_{12}) \left(A(k,\l) + B(k,\l) \ln {\varepsilon^2\ov |x_{12}|^2}\right) \ .
\end{split}
\end{equation}
At the conformal point this correlator should vanish. We have also allowed for
contact terms proportional to the $\d$-function since these are allowed by symmetry.
The coupling functions $A$ and $B$ have to be computed.

\no
After a long computation, all details
are given in the appendix \ref{detaJbarJ}, we found the result
\ba
&& \langle J^a(x_1)\bar J^b(\bar x_2)\rangle_\l =  -\pi \l \d^{ab} \d^{(2)}(x_{12})
\nonumber\\
&& \phantom{xxxx} - {\l^2 c_G\ov k}  \d^{ab} \left[{1\ov |x_{12}|^2}
+ \pi \d^{(2)}(x_{12})\left(1-\ha \ln {\varepsilon^2 \ov  |x_{12}|^2} \right)\right]
\label{JbJaiso}
\\
&& \phantom{xxxx} + 2{\l^3 c_G\ov k}  \d^{ab} \left[{1\ov |x_{12}|^2}
+ \pi \d^{(2)}(x_{12})\left(1- \ln {\varepsilon^2 \ov  |x_{12}|^2} \right)\right] + {1\ov k} {\cal O}(\l^4)\ .
\nonumber
\ea
which, keeping in mind that we are interested to terms up to ${\cal O}(1/k)$,
is easily seen to be of the form \eqn{generic.corr2}. Note that this correlator takes the form
\begin{equation}
\langle J^a(x_1)\bar J^b(\bar x_2)\rangle = -\gamma^{(J)} {\delta_{ab}\ov |x_{12}|^{ 2}} + \text{contact\ terms}\ ,
\end{equation}
where $\g^{(J)}$ is the current anomalous dimension given perturbatively by \eqn{gammapert1}.

\subsubsection*{The exact $\beta$-function and anomalous dimensions}

To compute the wave function renormalization and that for the parameter $\lambda$ we use the
two-point functions $\langle J^a J^b \rangle$ and $\langle J^a \bar J^b \rangle$.
In particular we need the most singular part of these correlation functions.
For the purpose of this section let's denote the bare currents by $J^a_0$ and
$\bar J^a_0$ and similarly for the parameter $\lambda_0$.

\no
We need the most singular part of the bare two-point functions up to order $1/k$.
From \eqn{jjll} we have that
\begin{equation}
\label{kfkfhhf}
\langle J^a_0(x_1) J^b_0(x_2)\rangle
=\frac{\delta_{ab}}{x_{12}^2}\Big[1 - {c_G\ov k}\lambda_0^2\left(2 \lambda_0
+ (1-2 \lambda_0) \ln(|x_{12}|^2/{ \varepsilon^2})\right)\Big] + \dots \ .
\end{equation}
Also from \eqn{JbJaiso} we have that
\be
\begin{split}
& \langle J_0^a(x_1)\bar J_0^b(\bar x_2)\rangle =  -\pi\lambda_0\, \d^{ab} \d^{(2)}(x_{12})
\Big[1 +\lambda_0 {c_G\ov k}\bigg(1-\ha \ln{\varepsilon^2\ov |x_{12}|^2}
\\
&\phantom{xxxx}
- 2\lambda_0 \left(1- \ln {\varepsilon^2\ov |x_{12}|^2} \right) \bigg) \Big] + \cdots\ ,
\end{split}
\ee
where we have kept only the coefficient of the most singular term, i.e. of $\d^{(2)}(x_{12})$.

\no
The bare quantities and the renormalized ones are related as
\begin{equation}
J_0^a =Z^{1/2} J^a \ ,\qq \bar J_0^a = Z^{1/2} \bar J^a\ , \qq \lambda_0= Z_1 \l\ .
\end{equation}
We make the following ansatz valid to order $1/k$ in the large $k$-expansion
\be
\begin{split}
&  Z^{-1} = 1+ 2 {c_G\ov k} \l^3 -  {c_G\ov k}\left(c_1 \l^2 + c_2 \l^3 + {\cal O}(\lambda^4) \right)\ln(\varepsilon^2 \m^2)\ ,
\\
& Z_1 = 1- {c_G\ov k}\left(c_3 \l + c_4 \l^2 + {\cal O}(\lambda^3) \right)\ln(\varepsilon^2 \m^2)\ ,
\end{split}
\ee
where the logarithm-independent term in $Z^{-1}$ has been chosen so that the renormalized two-point function for the $J_a$'s
is normalized to one.
The pure number coefficients $c_i$ are computed so that the renormalized two-point functions
\begin{equation}
\langle J^a(x_1) J^b(x_2)\rangle =Z^{-1} \langle J^a_0(x_1) J^b_0(x_2)\rangle\ ,\qq
\langle J^a(x_1) \bar J^b(x_2)\rangle =Z^{-1} \langle J^a_0(x_1) \bar J^b_0(x_2)\rangle\ ,
\end{equation}
are independent of the cutoff $\varepsilon$.
We find that the unique choice is given by
\begin{equation}
c_1= 1\ ,\qq c_2 = -2 \ ,\qq c_3 =-\frac{1}{2} \ ,\qq c_4 = 1\ .
\end{equation}
The $\b$-function is by definition
\begin{equation}
\b_\l =\ha \m {\text{d}\l\ov \text{d}\m} = \ha \l Z_1 \m {\text{d} Z_1^{-1}\ov \text{d}\m} = -{c_G\ov 2k} \left(\l^2 -2 \l^3 + {\cal O}(\l^4)\right)\ ,
\label{pertb}
\end{equation}
where the bare coupling coupling $\lambda_0$ is kept fixed.
Next we compute the anomalous dimension of the current
\begin{equation}
\gamma^{(J)}=\mu\frac{\text{d}\ln Z^{1/2}}{\text{d}\mu}={c_G\ov k} \left(\l^2 -2 \l^3 + {\cal O}(\l^4)\right)\ ,
\label{pertg}
\end{equation}
in agreement of course with \eqn{gammapert1}.

\no
The above perturbative expressions are enough to determine the exact in $\lambda$ dependence of the
$\b$-function and of the anomalous dimensions up to order $1/k$.
As explained, the exact $\b$-function and anomalous dimensions should have a well defined behaviour in the two limiting
cases described by the non-Abelian and pseudodual model limits \eqn{laborio} and \eqn{limps}, respectively.
In the isotropic case, which is the case of interest in this work, it implies regularity under the following
independent limits
\begin{equation}
\lambda= 1-{\kappa^2\ov k}\ ,\qq \lambda = - 1 + {1\ov b^{2/3} k^{1/3}}\ ,\qq k\to \infty\ .
\label{limnonps}
\end{equation}
{ Regularity under \eqref{limnonps} of the exact $\b$-function and the anomalous dimensions implies an ansatz of the form}
\begin{equation}
\b_\lambda= -{c_G\ov 2 k} {f(\lambda)\ov (1+\lambda)^2}\ ,\qq \g^{(J)} = {c_G\ov k} {g(\lambda)\ov (1-\lambda)(1+\lambda)^3}\ ,
\end{equation}
where $f(\l)$ and $g(\l)$ are two analytic functions of $\lambda$. The assumed pole structure does
not exclude the possibility that one of the poles reduces its degree or even {ceases} to exist.
This can happen if the functions in the numerator are zero at $\l=1$ or/and $\l=-1$.
In addition, due to the symmetry under $(k,\l)\mapsto (-k,\l^{-1})$ we have that
\begin{equation}
\lambda^4 f(1/\lambda)= f(\lambda)\ ,\qq \lambda^4 g(1/\lambda)= g(\lambda)\ .
\end{equation}
All these imply that these functions are in fact polynomials of, at most, degree four
\begin{equation}
\begin{split}
& f(\l)= a_0 + a_1 \l + a_2 \l^2 + a_1\l^3+ a_0 \l^4 \ ,
\\
& g(\l)= b_0 + b_1 \l + b_2 \l^2 + b_1\l^3+ b_0 \l^4\ .
\end{split}
\end{equation}
Demanding agreement with the perturbative expressions \eqn{pertb} and \eqn{pertg} to ${\cal O}(\l^2)$ we obtain
$a_0=a_1=b_0=b_1=0$ and $a_2=b_2=1$ which completely determines the exact $\b$-function and anomalous
dimensions to be
\begin{equation}
\boxed{\b_\l = -{c_G\ov 2 k} {\l^2\ov (1+\l)^2}\leqslant0}
\end{equation}
and
\begin{equation}\label{gamma-exact}
\boxed{\g^{(J)} = {c_G\ov k} {\l^2\ov  (1-\l)(1+\l)^3}\geqslant0.}
\end{equation}
It is also easily seen that the coefficient of the ${\cal O}(\l^3)$ term is in agreement with the
perturbative results as well. The above expressions are in full agreement with the results found in
\cite{Sfetsos:2014jfa,Itsios:2014lca,Appadu:2015nfa} for the $\beta$-function and in \cite{Georgiou:2015nka} for the anomalous dimensions.

\no
Note that the $\b$-function and anomalous dimensions of the non-Abelian T-dual limit are
\begin{equation}
\b_{\kappa^2} = {c_G\ov 8}\ ,\quad \g^{(J)} = {c_G\ov 8\k^2}\ ,
\end{equation}
which are valid for large $\kappa^2$.
The anomalous dimensions correspond to
\begin{equation}
J_\pm^a =\pm \ha (\kappa^2 \mathbb{I} \mp  f)^{-1}_{ab} \del_\pm v^b \ ,
\end{equation}
which are obtained by taking this limit in \eqn{jredf}.

\no
The corresponding expressions for the pseudodual model are
\begin{equation}
\b_{b} = {3\ov 4} c_G b^3\ ,\qq \g^{(J)} = \ha c_G b^2\ .
\end{equation}
These are in agreement with the expressions derived in \cite{Nappi:1979ig} (see above fig. 2)
and are valid for small $b$.
The anomalous dimensions correspond to
\begin{equation}
J_\pm^a = \pm b^{2/3}\del_\pm  v^a \ ,
\end{equation}
which as before are obtained by taking the appropriate limit in \eqn{jredf}.

\subsection{Three-point functions}\label{sec:jjj3}

We consider the $\langle JJJ\rangle$ and $\langle JJ\bar J\rangle$ correlators.
The remaining correlators $\langle \bar J\bar J\bar J\rangle$ and $\langle \bar J\bar J J\rangle$ can be easily obtained
by applying the parity transformation to the first two. The results of this subsection match those obtained in \cite{Konechny:2010nq},
where current-current perturbations of the WZW model on supergroups were studied with a different regularization scheme.
Before moving to our analysis, let us note that analogue perturbations of the WZW models on supergroups were studied in
\cite{Ashok:2009xx}, but the perturbation consists of the term $J^a_+ D_{ab} J^b_-$ added to the action; effectively the non-critical WZW model.

\subsubsection*{ The $\langle JJJ \rangle$ correlator}

From appendix \ref{jjjcor} we have that the, up to ${\cal O}(\l^3)$, correlator reads
\begin{equation}
\label{jfkfkksk}
\langle J^a(x_1) J^b(x_2) J^c(x_3)\rangle_\l
=\frac{1}{\sqrt{k}}\left(1+\frac32\,\lambda^2-\lambda^3\right)
\frac{f_{abc}}{x_{12}x_{13}x_{23}} +{1\ov \sqrt{k}} {\cal O}(\l^4) \ .
\end{equation}
The ansatz for the all-loop expression takes the form
\begin{equation}
\label{kflflk}
\langle J^a(x_1) J^b(x_2) J^c(x_3)\rangle=\frac{f(\l)}{\sqrt{k(1-\lambda)(1+\lambda)^3}} \frac{f_{abc}}{\,x_{12}x_{13}x_{23}}\,,
\end{equation}
where $f(\lambda)$ is everywhere analytic and obviously $f(0)=1$ to agree with the CFT result.
As before this form takes into account that under the limit \eqn{limnonps} the correlator is well behaved.
Invariance of the above expression under the duality-type symmetry $(k,\l)\mapsto (-k,\l^{-1})$ yields
\begin{equation}
\lambda^2f(\lambda^{-1})=f(\lambda) \quad \Longrightarrow\quad f(\lambda)=1+c\,\lambda+\lambda^2\ .
\end{equation}
Consistency with the perturbative expression up to ${\cal O}(\l)$ \eqref{kflflk} gives $c=1$.
Therefore, the all-loop correlator reads
\begin{equation}
\boxed{
\langle J^a(x_1) J^b(x_2) J^c(x_3)\rangle= {1+\l+ \l^2\ov \sqrt{k(1-\lambda)(1+\l)^3}}
\frac{f_{abc}}{x_{12}x_{13}x_{23}}} \ .
\end{equation}
As a check we see that this expression reproduces the ${\cal O}(\l^2)$ and ${\cal O}(\l^3)$
terms in the perturbative expression \eqn{jfkfkksk}.

\subsubsection*{ The $\langle JJ \bar J \rangle$ correlator}

The perturbative calculation of this correlator is performed in appendix \ref{jjbjcor}. The result up to order ${\cal O}(\l^2)$  reads
\begin{equation}
\label{mfjkdkdbn}
\langle J^a(x_1) J^b(x_2) \bar J^c(\bar x_3)\rangle
=\frac{\lambda(1-\lambda)}{\sqrt{k}}\,\frac{\bar x_{12}\,f_{abc}}{x_{12}^2\bar x_{23}\bar x_{13}}
+{1\ov \sqrt{k}} {\cal O}(\l^3) \ .
\end{equation}
We now make a similar to \eqn{kflflk} ansatz for the all-loop expression
\begin{equation}
\label{kflflkmm}
\langle J^a(x_1) J^b(x_2) \bar J^c(x_3)\rangle= \frac{\lambda f(\l)}{\sqrt{k(1-\lambda)(1+\lambda)^3}}
\frac{\bar x_{12}\,f_{abc}}{x_{12}^2\bar x_{23}\bar x_{13}}\ ,
\end{equation}
where $f(\lambda)$ is everywhere analytic and $f(0)=1$.
Invariance of the above expression under the duality-type symmetry yields
\begin{equation}
f(\lambda^{-1})=f(\lambda)\quad \Longrightarrow\quad  f(\lambda)=1\,.
\end{equation}
Hence, we find the all-loop expression
\begin{equation}
\boxed{
\langle J^a(x_1) J^b(x_2) \bar J^c(\bar x_3)\rangle= {\l \ov \sqrt{k(1-\l)(1+\l)^3}}
\frac{f_{abc}\bar x_{12}}{x^2_{12}\bar x_{13}\bar x_{23}} }\ ,
\end{equation}
whose expansion around $\lambda=0$ agrees with \eqref{mfjkdkdbn}.

\no
Note that implementing the non-Abelian and pseudodual limits both lead to finite (non-zero) expressions for
all of the above three-point functions. In these limiting cases the results are valid for large $\k^2$ and small $b$ where we
refer to \eqn{limnonps} for the definition of these parameters.
We mention also that, our results for these correlators agree with those done for supergroups in
\cite{Konechny:2010nq} after an appropriate rescaling of the currents that presumably takes into account the different regularization
schemes used in that work.

\section{Primary field correlators}

The purpose of this section is to compute two- and three- point functions of arbitrary primary fields. This will allow us to
extract their anomalous dimensions and the deformed structure constants in the OPEs.

\subsection{Two-point functions}\label{sec:prim2}

After a long computation, all details of which are given in the appendix \ref{ffcor}, we found
that a perturbative computation up to ${\cal O}(\l^3)$ and to order $1/k$, gives for
the two-point function of primary fields the result
\ba
&& \langle \Phi^{(1)}_{i,i'}(x_1,\bar x_1) \Phi^{(2)}_{j,j'}(x_2, \bar x_2)\rangle_\l =
{1\ov x_{12}^{2 \D_R} \bar x_{12}^{2 \bar \D_{R'}}}
\bigg[\left(1+{\l^2\ov k} (c_R + c_{R'}) \ln{\varepsilon^2\ov |x_{12}|^2}\right) (\mathbb{I}_R \otimes \mathbb{I}_{R'})_{ii',jj'}
\nonumber\\
&& \qq\qq\qq -2\l {1+\l^2\ov k}   \ln{\varepsilon^2\ov |x_{12}|^2} (t_a \otimes t_a^*)_{ii',jj'}\bigg] + {1\ov k}{\cal O}(\l^3)\ .
\label{corrlf1}
\ea
We see that due to the deformation there is an operator mixing so that one should proceed by choosing an appropriate
basis in which the dimension matrix is diagonal. For convenience we will adopt the double index notation
$I=(ii')$. Then there is a matrix $U$ chosen such that
\begin{equation}
(t_a \otimes t_a^*)_{IJ}= U_{IK} N_{KL} (U^{-1})_{LJ}\ ,\qquad N_{IJ}=N_I \delta_{IJ}\ ,
\end{equation}
where $N_I$ are the eigenvalues of the matrix $t_a \otimes t_a^*$.
Note also that $U$ is $\lambda$-independent as well as $k$-independent.
Then in the rotated basis
\begin{equation}
\widetilde \Phi^{(1)}_I = (U^{-1})_I{}^J \Phi^{(1)}_J \ ,\qq  \widetilde \Phi^{(2)}_I = U_I{}^J \Phi^{(2)}_J\ ,
\end{equation}
the correlator \eqn{corrlf1} becomes diagonal, i.e.
\begin{equation}
\langle \widetilde \Phi^{(1)}_I(x_1,\bar x_1) \widetilde \Phi^{(2)}_J(x_2, \bar x_2)\rangle_\l =
{\delta_{IJ} \ov x_{12}^{2 \D_R} \bar x_{12}^{2 \bar \D_{R'}}}
\left(1 + \d^{(\Phi)}_I \ln{\varepsilon^2\ov |x_{12}|^2} \right) \ ,
\end{equation}
where perturbatively
\begin{equation}
\d^{(\Phi)}_I = {1\ov k} \left(- 2 \l (1+\l^2) N_I + \l^2(c_R + c_{R'}) + {{\cal O}(\l^4)}\right)\ .
\label{corrl2}
\end{equation}
To determine the exact anomalous dimension of the general primary field we first realize that we should
include in the above expression the $k$-dependent part coming
from the CFT dimensions of $\D_R$ and $\bar \D_{R'}$ in \eqn{ddcft} up to order $1/k$.
Hence the anomalous dimension is given by
\begin{equation}
\begin{split}
&
\g^{(I)}_{R,R'}(k,\l)\big |_{\rm pert} = {c_R\ov 2 k} +  {\d^{(\Phi)}_I\ov 2}=
\\
& \quad = {1\ov 2k} \left[c_R - 2 N_I \l(1+\l^2)  +\l^2 (c_R + c_{R'}) +{{\cal O}(\l^4)}\right]\ .
\end{split}
\label{petff}
\end{equation}
As in the case of currents we make the following ansatz for the exact anomalous dimensions
\begin{equation}
 \g^{(I)}_{R,R'}(k,\l) = -{1\ov 2k (1-\l) (1+\l)^3}\left[ f(\l) N_I + f_1(\l) c_R + f_2(\l) c_{R'}\right]\ ,
\label{girrr}
\end{equation}
where the yet unknown function should be analytic in $\lambda$.
Using the symmetry \eqn{dualsym} and the transformation of the primary fields
under this symmetry \eqn{ffsym}, we have that
 \begin{equation}
\g^{(I)}_{R,R'}(-k,\l^{-1}) =  \g^{(I)}_{R',R}(k,\l)\ ,
\end{equation}
which implies the following relations between the various unknown functions
\begin{equation}
\l^4 f(1/\l) =f(\l)\ ,\qq \l^4 f_1(1/\l) =f_2(\l)\ ,\qq \l^4 f_2(1/\l) =f_1(\l)\ .
\end{equation}
Hence, these functions should be fourth order polynomials in $\lambda$ with related coefficients.
It turns out that comparing with the perturbative expression \eqn{petff} up to ${\cal O}(\l^2)$ we determine
all these functions to be
\begin{equation}
f(\l) = 2\l (1+\l)^2\ ,\qq f_1(\l) = -(1+\l)^2\ , \qq f_2(\l) = -\l^2(1+\l)^2\ .
\end{equation}
Therefore, the exact in $\lambda$ anomalous dimension is
\begin{equation}
\boxed{ \g^{(I)}_{R,R'}(k,\l) = -{1\ov 2k (1-\l^2)}( 2 \l N_I  - c_R - \l^2 c_{R'})}\ .
\end{equation}
It is easily checked that this expression is in agreement with the ${\cal O}(\l^3/k)$ term
in \eqn{petff}.
Note also that in the non-Abelian limit the above anomalous dimensions have
a well defined and different than
zero limit. In contrast the limit is zero in the pseudodual limit.
{
This expression also applies for current
current perturbations of the WZW model on supergroups with vanishing Killing form  \cite{Candu:2012xc}.}

\no
Finally, the two point functions take the form
\begin{equation}
\boxed{\langle \Phi^{(1)}_{I}(x_1,\bar x_1) \Phi^{(2)}_{J}(x_2,\bar x_2)\rangle =  {\delta_{IJ}\ov
x_{12}^{\g^{(I)}_{R,R'}(k,\l)} \bar x_{12}^{\g^{(I)}_{R',R}(k,\l)} }\ .}
\end{equation}

\subsection{Three-point functions}\label{sec:prim3}

To leading order in the $\lambda$-expansion after a straightforward computation this correlator is found to be
\ba
&& \langle \Phi^{(1)}_{i,i'}(x_1) \Phi^{(2)}_{j,j'}(x_1) \Phi^{(3)}_{k,k'}(x_3)\rangle^{(1)}_\l =  -{\l\ov k} {1\ov
x_{12}^{\D_{12;3}}  x_{13}^{\D_{13;2}} x_{23}^{\D_{23;1}} \bar x_{12}^{\bar\D_{12;3}}  \bar x_{13}^{\bar\D_{13;2}}
\bar x_{23}^{\bar \D_{23;1}}}
\nonumber \\
&& \qq \Bigg[\ln \varepsilon^2 \left( (t_a^{(1)})_i{}^\ell (\tilde t_a^{(1)}){}^{\ell'}{}_{i'} C_{\ell \ell',jj',kk'} + (t_a^{(2)})_j{}^\ell (\tilde t_a^{(2)}){}^{\ell'}{}_{j'} C_{ii',\ell \ell',kk'} +
(t_a^{(3)})_k{}^\ell (\tilde t_a^{(3)}){}^{\ell'}{}_{k'} C_{ii',jj',\ell\ell'}
\right)
\nonumber \\
&&\qq  + \ln |x_{12}|^2 \left( (t_a^{(1)})_i{}^\ell (\tilde t_a^{(2)})^{\ell'}{}_{j'} C_{\ell i',j \ell',kk'}
+ (t_a^{(2)})_j{}^\ell (\tilde t^{(1)})^{\ell'}{}_{i'} C_{i \ell'\ell j',kk'}\right)
\\
&&
\qq + \ln |x_{13}|^2 \left( (t_a^{(1)})_i{}^\ell (\tilde t_a^{(3)})^{\ell'}{}_{k'} C_{\ell i',j j',k\ell'}
+ (t_a^{(3)})_k{}^\ell (\tilde t^{(1)})^{\ell'}{}_{i'} C_{i \ell',jj',\ell k'}\right)
\nonumber\\
&&
\qq +  \ln |x_{23}|^2 \left( (t_a^{(2)})_j{}^\ell (\tilde t_a^{(3)})^{\ell'}{}_{k'} C_{ii',\ell i',k\ell'}
+ (t_a^{(3)})_k{}^\ell (\tilde t^{(2)})^{\ell'}{}_{j'} C_{i i',j\ell',\ell k'}\right)\bigg]\ .
\nonumber
\ea
Even for dimensional reasons we should be able to cast the above expression in a
form in which all space dependence is in terms of ratios $\varepsilon^2/|x_{ij}|^2$.
In order to do that we first recall that the structure constants $C_{ii',jj',kk'}$
are factorized according to their {holomorphic} and antiholomorphic content as
\begin{equation}
C_{ii',jj',kk'} = C_{i,j,k}\tilde C_{i',j',k'}\ .
\end{equation}
An important constraint, arises by making use of the global Ward identity.
It reads
\begin{equation}
\label{ttrr}
\begin{split}
& (t^{(1)}_a)_i{}^\ell C_{\ell j k } + (t^{(2)}_a)_j{}^\ell C_{i\ell k } + (t^{(3)}_a)_k{}^\ell C_{i j \ell } = 0 \ ,
\\
& (\tilde t^{(1)}_a)^{\ell'}{}_{i'}{}\tilde C_{\ell' j' k' } + (\tilde t^{(2)}_a)^{\ell'}{}_{j'}{}\tilde C_{i'\ell'  k' }
+ (\tilde t^{(3)}_a)^{\ell'}{}_{k'}{}\tilde C_{i' j'\ell' }= 0\ .
\end{split}
\end{equation}
From \eqref{ttrr} it is straightforward to obtain the following relations
\ba
&& (t^{(3)}_a)_k{}^\ell (\tilde t^{(3)}_a)^{\ell'}{}_{k'} C_{ii',jj',\ell\ell'}= (t_a^{(1)})_i{}^\ell (\tilde t^{(1)}_a)^{\ell'}{}_{i'} C_{\ell \ell',jj',kk'} +
(t_a^{(2)})_j{}^\ell (\tilde t^{(2)}_a)^{\ell'}{}_{j'} C_{ii',\ell \ell',kk'}
\nonumber\\
 &&
\qq\qq \phantom{xxxxx}
+ (t_a^{(1)})_i{}^\ell (\tilde t^{(2)}_a)^{\ell'}{}_{j'} C_{\ell i',j \ell',kk'}
+ (t_a^{(2)})_j{}^\ell (\tilde t^{(1)}_a)^{\ell'}{}_{i'} C_{i\ell',\ell j',kk'}\ ,
\nonumber\\
&& (t^{(2)}_a)_j{}^\ell (\tilde t^{(2)}_a)^{\ell'}{}_{j'} C_{ii',\ell\ell',kk'}=
(t_a^{(1)})_i{}^\ell (\tilde t^{(1)}_a)^{\ell'}{}_{i'} C_{\ell \ell',jj',kk'} +
(t_a^{(3)})_k{}^\ell (\tilde t^{(3)}_a)^{\ell'}{}_{k'} C_{ii',jj',\ell \ell'}
\nonumber\\
 &&
\qq\qq \phantom{xxxxx}
+ (t_a^{(1)})_i{}^\ell (\tilde t^{(3)}_a)^{\ell'}{}_{k'} C_{\ell i',j j',k\ell'}
+ (t_a^{(3)})_k{}^\ell (\tilde t^{(1)}_a)^{\ell'}{}_{i'} C_{i\ell',jj',\ell k'}\ ,
\nonumber
\\
&& (t^{(1)}_a)_i{}^\ell (\tilde t^{(1)}_a)^{\ell'}{}_{i'} C_{\ell\ell',jj',kk'}=
(t_a^{(2)})_j{}^\ell (\tilde t^{(2)}_a)^{\ell'}{}_{j'} C_{ii',\ell \ell',kk'} +
(t_a^{(3)})_k{}^\ell (\tilde t^{(3)}_a)^{\ell'}{}_{k'} C_{ii',jj',\ell \ell'}\ ,
\nonumber\\
 &&
\qq\qq \phantom{xxxxx}
+ (t_a^{(2)})_j{}^\ell (\tilde t^{(3)}_a)^{\ell'}{}_{k'} C_{ii',\ell j',k\ell'}
+ (t_a^{(3)})_k{}^\ell (\tilde t^{(2)}_a)^{\ell'}{}_{j'} C_{ii',j\ell',\ell k'}\ .
\nonumber
\ea
Using the above relations we can rewrite the three-point function as
\ba
&& \langle \Phi^{(1)}_{i,i'}(x_1) \Phi^{(2)}_{j,j'}(x_1) \Phi^{(3)}_{k,k'}(x_3)\rangle^{(1)}_\l =  -{\l\ov k} {1\ov
x_{12}^{\D_{12;3}}  x_{13}^{\D_{13;2}} x_{23}^{\D_{23;1}} \bar x_{12}^{\bar\D_{12;3}}  \bar x_{13}^{\bar\D_{13;2}}
\bar x_{23}^{\bar \D_{23;1}}}
\nonumber \\
&& \hskip -.7 cm \Bigg[\ln \left(\varepsilon^2\ov |x_{12}|^2\right) \left[ (t_a^{(1)})_i{}^\ell (\tilde t_a^{(1)}){}^{\ell'}{}_{i'} C_{\ell \ell',jj',kk'}
+ (t_a^{(2)})_j{}^\ell (\tilde t_a^{(2)}){}^{\ell'}{}_{j'} C_{ii',\ell \ell',kk'}
-(t_a^{(3)})_k{}^\ell (\tilde t_a^{(3)}){}^{\ell'}{}_{k'} C_{ii',jj',\ell\ell'}\right]
\nonumber\\\
&& \hskip -.7 cm \ln \left(\varepsilon^2\ov |x_{13}|^2\right) \left[ (t_a^{(1)})_i{}^\ell (\tilde t_a^{(1)}){}^{\ell'}{}_{i'} C_{\ell \ell',jj',kk'}
+ (t_a^{(3)})_k{}^\ell (\tilde t_a^{(3)}){}^{\ell'}{}_{k'} C_{ii',jj',kk'}
-(t_a^{(2)})_j{}^\ell (\tilde t_a^{(2)}){}^{\ell'}{}_{j'} C_{ii',\ell\ell',kk'}\right]
\nonumber\\
&& \hskip -.7 cm \ln \left(\varepsilon^2\ov |x_{23}|^2\right) \left[ (t_a^{(2)})_j{}^\ell (\tilde t_a^{(2)}){}^{\ell'}{}_{j'} C_{ii',\ell \ell',kk'}
+ (t_a^{(3)})_k{}^\ell (\tilde t_a^{(3)}){}^{\ell'}{}_{k'} C_{ii',jj',kk'}
-(t_a^{(1)})_i{}^\ell (\tilde t_a^{(1)}){}^{\ell'}{}_{i'} C_{\ell\ell',jj',kk'}\right]
\Biggr]\ .
\nonumber
\ea
The next step is to pass to the rotated basis. By using the double index notation we introduced before
we have that
\be
\begin{split}
&
\tilde \Phi^{(q)}_I = (U^{(q)})^{-1}_I{}^J \Phi^{(q)}_J\ ,\quad (t_a^{(q)}\otimes t_a^{(q)*})_I{}^J =
(U^{(q)})_I{}^K (N^{(q)})_K{}^L (U^{(q)})^{-1}_L{}^J\ ,
\\
&
N^{(q)}_{IJ} = N_I^{(q)} \delta_{IJ} \ ,\quad q=1,2,3\ ,
\end{split}
\ee
where in the new basis the structure constants read
\begin{equation}
\tilde C_{IJK} = (U^{(1)})^{-1}_I{}^M (U^{(2)})^{-1}_J{}^N (U^{(3)})^{-1}_K{}^L C_{MNL}\ ,
\end{equation}
 while the result for the correlator at ${\cal O}(\l)$ is given by
\ba
&& \langle \tilde\Phi^{(1)}_{I}(x_1) \tilde\Phi^{(2)}_{J}(x_1) \tilde\Phi^{(3)}_{K}(x_3)\rangle^{(1)}_\l
=
-{\l\ov k} {\tilde C_{IJK}\ov
x_{12}^{\D_{12;3}}  x_{13}^{\D_{13;2}} x_{23}^{\D_{23;1}} \bar x_{12}^{\bar\D_{12;3}}  \bar x_{13}^{\bar\D_{13;2}}
\bar x_{23}^{\bar \D_{23;1}}}
\nonumber
\\
&& \qq \qq \left((N^{(1)}_I + N^{(2)}_I - N^{(3)}_I) \ln{\varepsilon^2\ov |x_{12}|^2} + \text{cyclic in 1,2,3}  \right)\ .
\ea
From this result we can write down the exact expression in $\lambda$ for the three-point function. It is given by
\begin{equation}
\boxed{
\langle \tilde\Phi^{(1)}_{I}(x_1) \tilde\Phi^{(2)}_{J}(x_1) \tilde\Phi^{(3)}_{K}(x_3)\rangle_\l
=  {\tilde C_{IJK}(k,\l)\ov
x_{12}^{\g_{12;3}/2}  x_{13}^{\g_{13;2}/2} x_{23}^{\g_{23;1}/2} \bar x_{12}^{\bar\g_{12;3}/2}  \bar x_{13}^{\bar\g_{13;2}/2}
\bar x_{23}^{\bar \g_{23;1}/2}}\ ,}
\end{equation}
where $\g_{12;3}$ is given by
\begin{equation}
\label{g123}
\begin{split}
\g_{12;3}   = & -{1\ov 2k (1-\l^2)}\Big(  2 \l (N^{(1)}_I+ N^{(2)}_I - N^{(3)}_I )
\\
&  - c_{R_1} - c_{R_2} + c_{R_3} - \l^2 (c_{R'_1}+ c_{R'_2} - c_{R'_3})\Big)\ .
\end{split}
\end{equation}
and
\begin{equation}
\label{g123p}
\begin{split}
\bar \g_{12;3} & = -{1\ov 2k (1-\l^2)}\Big( 2 \l (N^{(1)}_I+ N^{(2)}_I - N^{(3)}_I )
\\
& - c_{R'_1} - c_{R'_2} + c_{R'_3}- \l^2 ( c_{R_1} + c_{R_2}- c_{R_3})\Big)\ .
\end{split}
\end{equation}
The other differences of dimensions $\g_{23;1},\bar \g_{23;1}\, $ and $\g_{13;2},\bar \g_{13;2}\, $ are obtained by performing  cyclic permutations in the indices $1,2$ and $3$.

We now turn our attention to  the three-point function coefficients $\tilde C_{IJK}(k,\l)$.
At $\l=0$ these coefficients are considered as known since they are in principle fully determined from the WZW CFT
data.
On general grounds the following perturbative expansion holds
\begin{equation}
\label{tildeCs}
\tilde C_{IJK}(k,\l)= \tilde C_{IJK}^{(0)} + {1\ov k}\tilde C_{IJK}^{(1)}(\l)+\mathcal{O}\left({1 \ov k^2}\right)\ .
\end{equation}
where note the leading coefficient $\tilde C_{IJK}^{(0)}$ in $1/k$ expansion does not depend on $\lambda$.
This is so because such a term being $k$-independent and simultaneously having possible poles only at $\l=\pm 1$
and preserving the symmetry $k\mapsto -k, \lambda\mapsto \lambda^{-1}$ cannot be finite  either in the non-Abelian T-dual or in
the pseudodual limit.
Using the same line of reasoning as in the rest of this paper we conclude that the first correction to the three-point function should be
of the form\footnote{Notice that here we are using the duality \eqref{dualsym} followed by parity. Under this combined symmetry
$\Phi^{(i)}_{I}(x_i,\bar x_i) \mapsto \Phi^{(i)}_{I}(\bar x_i, x_i)$ and $\tilde C_{IJK}(\l^{-1},-k)=\tilde C_{IJK}(\l,k)$.}
\begin{equation}
\tilde C_{IJK}^{(1)}(\l)={f_{IJK}(\l) \ov (1-\l)(1+\l)^3}\ ,
\end{equation}
with
\be
 \l^4\,f_{IJK}(\l^{-1})=f_{IJK}(\l)\  \Longrightarrow \ f_{IJK}(\l)
 =\tilde C_{IJK}^{(1)}(0)(1+ \l^4) +a_{IJK}^{(1)} (\l+\l^3) +a_{IJK}^{(2)} \l^2\ .
\ee
We saw from the ${\cal O}(\lambda)$ calculation that $a_{IJK}^{(1)}=0$.
Furthermore, it is not difficult to see that $a_{IJK}^{(2)}=0$ too.
Indeed, by inspecting the ${\cal O}(\lambda^2)$ calculation one can see that in order to remain to order $1/k$
either the two holomorphic or the two anti-holomorphic currents
should be contracted through the Abelian part of their OPE. Then the resulting integrals will be of the form
$\displaystyle \int {\text{d}^2z_{12} \ov (z_1-x_1)(z_2-x_2)\,\bar z^2_{12}}$ which can only  produce logarithms.
But the logarithms have to be combined and exponentiated to give
the differences of the anomalous dimensions. Thus, no finite part will be present
at this order and as a result $a_{IJK}^{(2)}=0$, as well.
Thus, we conclude that
\begin{equation}
\tilde C_{IJK}^{(1)}(\l)={\tilde C_{IJK}^{(1)}(0)(1+ \l^4) \ov (1-\l)(1+\l)^3}\ ,
\end{equation}
where as explained, the constant $\tilde C_{IJK}^{(1)}(0)$ is fully determined from the WZW CFT initial data.
As a result we have determined the exact in $\lambda$ three-point function coefficient of three-primary fields up to
order $1/k$.

\section{Mixed $\langle J \Phi \Phi\rangle$ and $\langle \bar J \Phi \Phi\rangle$ correlators}\label{sec:mixed}

In this section we focus on the mixed correlators involving two primary fields and one current.
From appendix \ref{jff} one can read off the ${\cal O}(\l^3)$ result which is given by
\begin{equation}
\label{jfkkfkf}
\begin{split}
&\langle J^a(x_3) \Phi_{i,i'}^{(1)}(x_1)\Phi^{(2)}_{j,j'}(x_2)\rangle_\l=\\
&\left(1+\frac{\lambda^2}{2}\right)
\frac{(t_a\otimes\mathbb{I}_{R'})_{ii',jj'}-\lambda\,(\mathbb{I}_{R}\otimes\tilde t^*_a)_{ii',jj'}}
{\sqrt{k}\,x_{12}^{2\Delta_R}\bar x_{12}^{2\bar\Delta_{R'}}}
\left(\frac{1}{ x_{13}}-\frac{1}{ x_{23}}\right) \ .
\end{split}
\end{equation}
The similar expression for the correlator $\bar J^a$ reads
\begin{equation}
\label{jfkkfkfa}
\begin{split}
&\langle \bar J^a(\bar x_3) \Phi_{i,i'}^{(1)}(x_1)\Phi^{(2)}_{j,j'}(x_2)\rangle_\l=\\
&- \left(1+\frac{\lambda^2}{2}\right)
\frac{(\mathbb{I}_{R}\otimes\tilde t^*_a)_{ii',jj'}-\lambda\,(t_a\otimes\mathbb{I}_{R'})_{ii',jj'}}
{\sqrt{k}\,x_{12}^{2\Delta_R}\bar x_{12}^{2\bar\Delta_{R'}}}
\left(\frac{1}{\bar x_{13}}-\frac{1}{\bar x_{23}}\right)\ .
\end{split}
\end{equation}
Getting inspired by the previous computations and by the expression in \eqref{jfkkfkf} we conclude that the all-loop
mixed correlators should assume the following form
\begin{equation}
\label{mfeldfmjs}
\begin{split}
&\langle J_a(x_3) \Phi_{i,i'}^{(1)}(x_1)\Phi^{(2)}_{j,j'}(x_2)\rangle_\l=\\
& \frac{f_1(\l)(t_a\otimes\mathbb{I}_{R'})_{ii',jj'}-\l f_2(\l)
(\mathbb{I}_{R}\otimes\tilde t^*_a)_{ii',jj'}}
{\sqrt{k(1-\lambda)(1+\lambda)^3}\,x_{12}^{2\Delta_R}
\bar x_{12}^{2\bar\Delta_{R'}}}\left(\frac{1}{ x_{13}}-\frac{1}{ x_{23}}\right)\ ,
\end{split}
\end{equation}
where the functions $f_1(\l)$ and $f_2(\lambda)$ are everywhere analytic and
$f_1(0)= f_2(0)=1$.
As usual, the denominator of \eqref{mfeldfmjs} is written in such a way that the correlator has well-defined non-Abelian and pseusodual limits.

\no
Applying  the duality \eqref{dualsym}, as well as the corresponding transformation rules for the currents \eqref{jjgh} and primary fields
 \eqref{ffsym} we obtain that
\begin{equation}\label{1st}
\begin{split}
&\langle J_a(x_3) \Phi_{i',i}^{(2)}(x_1)\Phi^{(1)}_{j',j}(x_2)\rangle_\l=\\
& \frac{\l^2 f_1(\l^{-1})(\mathbb{I}_{R'}\otimes t_a)_{i'i,j'j}
-\l f_2(\l^{-1})(\tilde t^*_a\otimes \mathbb{I}_{R})_{i'i,j'j}}
{\sqrt{k(1-\lambda)(1+\lambda)^3}\,x_{12}^{2\Delta_R}
\bar x_{12}^{2\bar\Delta_{R'}}}\left(\frac{1}{ x_{13}}-\frac{1}{ x_{23}}\right)\ ,
\end{split}
\end{equation}
where on the right hand side of the last equation we have changed the order of the indices for convenience.
Subsequently, the left hand side of the above can be rewritten using appropriately \eqn{mfeldfmjs}. We have that
\ba\label{2nd}
&& \langle J_a(x_3) \Phi_{i',i}^{(1)}(x_1)\Phi^{(2)}_{j',j}(x_2)\rangle_\l
= \frac{f_1(\l)(\tilde t^{(2)}_a\otimes\mathbb{I}_{R})_{i'i,j'j}
-\l f_2(\l)(\mathbb{I}_{R'}\otimes t^{(2)*}_a)_{i'i,j'j}}
{\sqrt{k(1-\lambda)(1+\lambda)^3}\,x_{12}^{2\Delta_R}\bar x_{12}^{2\bar\Delta_{R'}}}
\left(\frac{1}{ x_{13}}-\frac{1}{ x_{23}}\right)
\nonumber\\
&&\qq\qq\qq  = \frac{-f_1(\l)(\tilde t^*_a\otimes\mathbb{I}_{R})_{i'i,j'j}
 + \l f_2(\l)(\mathbb{I}_{R'}\otimes t_a)_{i'i,j'j}}
{\sqrt{k(1-\lambda)(1+\lambda)^3}\,x_{12}^{2\Delta_R}\bar x_{12}^{2\bar\Delta_{R'}}}
\left(\frac{1}{ x_{13}}-\frac{1}{ x_{23}}\right)\ .
\ea
Hence, comparing \eqref{1st} with \eqref{2nd} we have the two equivalent conditions
\begin{equation}
\lambda f_1(\lambda^{-1})=f_2(\lambda)\ ,\quad \lambda f_2(\lambda^{-1})=f_1(\lambda)\quad
\Longrightarrow \quad  f_1(\l)= f_2(\lambda)=1+\lambda\ .
\end{equation}
Plugging the latter into \eqref{mfeldfmjs} we find after some rearrangement that
\begin{equation}
\boxed{
\langle J^a(x_3)\Phi_{i,i'}^{(1)}(x_1,\bar x_1)  \Phi_{j,j'}^{(2)}(x_2,\bar x_2)\rangle_\l= -
{(t_a\otimes\mathbb{I}_{R'})_{ii',jj'}-\lambda (\mathbb{I}_{R}\otimes\tilde t^*_a)_{ii',jj'}\ov
\sqrt{k(1-\lambda^2)}  x_{12}^{2\Delta_R-1}\bar x_{12}^{2\bar\Delta_{R'}} x_{13} x_{23}}
\,.}
\end{equation}
Similar reasoning leads to
\begin{equation}
\boxed{
\langle \bar J^a(\bar x_3)\Phi_{i,i'}^{(1)}(x_1,\bar x_1)  \Phi_{j,j'}^{(2)}(x_2,\bar x_2)\rangle_\l
=
\frac{(\mathbb{I}_{R}\otimes\tilde t^*_a)_{ii',jj'}-\lambda(t_a\otimes\mathbb{I}_{R'})_{ii',jj'}}
{\sqrt{k(1-\lambda^2)}\,x_{12}^{2\Delta_R}\bar x_{12}^{2\bar\Delta_{R'}-1}  \bar x_{13} \bar x_{23}}
\,.}
\end{equation}
whose expansions around $\lambda=0$ {agree} with \eqref{jfkkfkf} and \eqref{jfkkfkfa}.
We stress that the one- and two-loop calculations
in conjunction with the symmetry \eqref{dualsym}
are enough to fully determine the all-loop expressions for the correlators under consideration.
Thus, the ${\cal O}(\l^3)$ terms in \eqref{jfkkfkf} and
\eqref{jfkkfkfa} provide perturbative checks of the all-loop results. Note that the deformation mixes
the left and right representations. It can be easily checked that the $\l$-deformed direct products
in the numerators in the above correlators form representations of the algebra as well.

\section{OPEs and equal-time commutators}\label{sec:OPE}

In this section we use the two-point and three-point
functions of the currents and primary fields to find their
OPE algebra up to order $1/k$ reads and exact in the deformation parameter $\l$.
The result is
\begin{equation}
\label{OPEfull}
\begin{split}
&J^a(x_1)J^b(x_2)=\frac{\delta_{ab}}{x_{12}^{2+\gamma^{(J)}}\bar x_{12}^{\gamma^{(J)}}}+c(\lambda)\frac{f_{abc}J^c(x_2)}{x_{12}}
+d(\lambda)\frac{f_{abc}\bar J^c(\bar x_2)\bar x_{12}}{x^2_{12}}+\dots\,,\\
&J^a(x_1)\bar J^b(\bar x_2)=-\gamma^{(J)} {\delta_{ab}\ov |x_{12}|^{ 2}}+d(\lambda)\frac{f_{abc}\bar J^c(\bar x_2)}{x_{12}}
+d(\lambda)\frac{f_{abc} J^c(x_2)}{\bar x_{12}}+\dots\,,\\
&J^a(x_1) \Phi_{i,i'}^{(1)}(x_2,\bar x_2)=-
\frac{(t_a)_i{}^m\Phi^{(1)}_{m,i'}(x_2,\bar x_2)-\lambda(\tilde t_a^*)_{i'}{}^{m'}\Phi^{(1)}_{i,m'}(x_2,\bar x_2)}
{x_{12}\sqrt{k(1-\lambda^2)}}+\dots\,,\\
&\bar J^a(\bar x_1) \Phi_{i,i'}^{(1)}(x_2,\bar x_2)=
\frac{(\tilde t_a^*)_{i'}{}^{m'}\Phi^{(1)}_{i,m'}(x_2,\bar x_2)-\lambda (t_a)_i^{m}\Phi^{(1)}_{m,i'}(x_2,\bar x_2)}
{\bar x_{12}\sqrt{k(1-\lambda^2)}}+\dots\,,\\
&\tilde\Phi_{I}^{(1)}(x_1,\bar x_1)\tilde\Phi_{J}^{(2)}(x_2,\bar x_2)=
\frac{\tilde C_{IJK}\, \tilde\Phi_{K}^{(3)}(x_2,\bar x_2)}{x_{12}^{\gamma_{12;3}}
\bar x_{12}^{\bar\gamma_{12;3}}}+\dots\,,\
\end{split}
\end{equation}
where $\tilde C_{IJK}$ was given in \eqref{tildeCs},
$\gamma^{(J)}$ is the anomalous dimension of the current given in \eqref{gamma-exact},
the $\g_{12;3}$ and $\bar \g_{12;3}$ are given by \eqn{g123} and \eqn{g123p}
and
\begin{equation}
 c(\lambda)=\sqrt{\frac{(1-\lambda^3)^2}{k(1-\lambda^2)^3}}\,,\qquad
d(\lambda)=\sqrt{\frac{\lambda^2(1-\lambda)^2}{k(1-\lambda^2)^3}}\,.
\end{equation}

Having the OPEs at our disposal, we can easily compute the equal-time commutator of the currents and primaries
through a time-ordered limiting procedure
\begin{equation}
\label{commut.lim}
[f(\sigma_1,\tau),g(\sigma_2,\tau)]=\lim_{\varepsilon\to0}
\left(f(\sigma_1,\tau + i\varepsilon)g(\sigma_2,\tau)-g(\sigma_2,\tau+i\varepsilon)f(\sigma_1,\tau)\right)
\end{equation}
and the following representations of Dirac delta-function
\begin{equation}
\label{Dirac}
\begin{split}
&\lim_{\varepsilon\to0}\left(\frac{1}{\sigma-i\varepsilon}
-\frac{1}{\sigma+i\varepsilon}\right)=2\pi i\,\delta(\sigma)\,,
\\
&\lim_{\varepsilon\to0}\left(\frac{1}{(\sigma-i\varepsilon)^2}
-\frac{1}{(\sigma+i\varepsilon)^2}\right)=-2\pi i\,\delta'(\sigma)\,,\\
&\lim_{\varepsilon\to0}\left(\frac{\sigma+i\varepsilon}{(\sigma-i\varepsilon)^2}-
\frac{\sigma-i\varepsilon}{(\sigma+i\varepsilon)^2}\right)=2\pi i\,\delta(\sigma)\,.
\end{split}
\end{equation}
Employing Eqs. \eqref{OPEfull}, \eqref{commut.lim} and \eqref{Dirac},
we find to order $\nicefrac{1}{\sqrt{k}}$ that\footnote{The OPEs and the equal-time commutators
for the currents are in agreement with
those obtained in \cite{Konechny:2010nq}, for current-current perturbations of the WZW model on supergroups.}
\begin{equation}
\begin{split}
\label{current.commut}
&[J^a(\sigma_1),J^b(\sigma_2)]=2\pi\, i\,\delta_{ab}\delta'(\sigma_{12})+
2\pi\,f_{abc}\left(\,c(\lambda)J^c(\sigma_2)-d(\lambda)\bar J^c(\sigma_2)\right)
\delta(\sigma_{12})\,,\\
&[\bar J^a(\sigma_1),\bar J^b(\sigma_2)]=-2\pi\, i\,\delta_{ab}\delta'(\sigma_{12})+
2\pi\,f_{abc}\left(c(\lambda)\bar J^c(\sigma_2)-d(\lambda)J^c(\sigma_2)\right)
\delta(\sigma_{12})\,,\\
&[J^a(\sigma_1),\bar J^b(\sigma_2)]=
2\pi\,d(\lambda)\,f_{abc}\left(J^c(\sigma_2)+\bar J^c(\sigma_2)\right)\delta(\sigma_{12})\,,
\end{split}
\end{equation}
and
\begin{eqnarray}
&&[\tilde\Phi^{(1)}_{i,i'}(\sigma_1),\tilde\Phi^{(2)}_{j,j'}(\sigma_2)]=0\,,\\
&&[J^a(\sigma_1),\Phi^{(1)}_{i,i'}(\sigma_2)]=-\frac{2\pi}{\sqrt{k(1-\lambda^2)}}\,
\left((t_a)_i{}^m\Phi^{(1)}_{m,i'}(\sigma_2)-\lambda(\tilde t_a^*)_{i'}{}^{m'}\Phi^{(1)}_{i,m'}(\sigma_2)\right)\,\delta(\sigma_{12})\,,\nonumber\\
&&[\bar J^a(\sigma_1),\Phi^{(1)}_{i,i'}(\sigma_2)]=\frac{2\pi}{\sqrt{k(1-\lambda^2)}}\,
\left((\tilde t_a^*)_{i'}{}^{m'}\Phi^{(1)}_{i,m'}(\sigma_2)-\lambda (t_a)_i{}^{m}\Phi^{(1)}_{m,i'}(\sigma_2)\right)\,\delta(\sigma_{12})\,.\nonumber
\end{eqnarray}
These equal-time commutators turn out to be isomorphic to two commuting copies of current algebras
with opposite levels
\be
\label{ssss}
\begin{split}
&[S^a(\sigma_1),S^b(\sigma_2)]=\frac{i\,k}{2\pi}\,\delta_{ab}\delta'(\sigma_{12})+
f_{abc}\,S^c(\sigma_2)\,\delta(\sigma_{12})\,,\\
&[\bar S^a(\sigma_1),\bar S^b(\sigma_2)]=-\frac{i\,k}{2\pi}\,\delta_{ab}\delta'(\sigma_{12})+
f_{abc}\,\bar S^c(\sigma_2)\,\delta(\sigma_{12})\,,\\
&[S^a(\sigma_1),\bar S^b(\sigma_2)]=0\,,\\
&[S^a(\sigma_1),\Phi^{(1)}_{i,i'}(\sigma_2)]=-(t_a)_i{}^m\Phi^{(1)}_{m,i'}(\sigma_2)\,\delta(\sigma_{12})\,,\\
&[\bar S^a(\sigma_1),\Phi^{(1)}_{i,i'}(\sigma_2)]=(\tilde t_a^*)_{i'}{}^{m'}\Phi^{(1)}_{i,m'}(\sigma_2)\,\delta(\sigma_{12})\,,
\end{split}
\ee
where
\be
\label{jkfllfld}
S^a=\frac{1}{2\pi}\sqrt{\frac{k}{1-\lambda^2}}\,\left(J^a-\lambda \bar J^a\right)\,,\quad
\bar S^a=\frac{1}{2\pi}\sqrt{\frac{k}{1-\lambda^2}}\,\left(\bar J^a-\lambda J^a\right)\,.
\ee
The parameter $\l$ does not appear in this algebra but it does in the time evolution of the
system due to the fact that, as it turns out, the Hamiltonian in terms of
$S^a$ and $\bar S^a$ is $\l$-dependent (cf. eq. (2.11) of \cite{Sfetsos:2013wia}).
Also, the $\l$-dependence still appears in the OPEs of the $S^a$ and $\bar S^a$ among them.
The reasons is that the OPEs, unlike the commutators \eqn{ssss}, are not computed at equal times.

\no
Finally we take the classical limit of \eqref{current.commut} and appropriately rescaling the currents, we find
Rajeev's {\it deformation of the canonical structure of the isotropic PCM}
\cite{Rajeev:1988hq} (recall that, in our conventions the group structure constants $f_{abc}$ are taken to be imaginary)
\begin{equation}
\boxed{
\label{defpoira}
\begin{split}
&\{I_\pm^a,I_\pm^b\}_{\text P.B.}=-i\,e^2f_{abc}
\left(I_\mp^c(\sigma_2)-(1+2x)I_\pm^c(\sigma_2)\right)\delta(\sigma_{12})\pm2e^2\delta_{ab}\,\delta'(\sigma_{12})\,,\\
&\{I_\pm^a,I_\mp^b\}_{\text P.B.}=i\,e^2f_{abc}
\left(I_+^c(\sigma_2)+I_-^c(\sigma_2)\right)\delta(\sigma_{12})\,,
\end{split}
}
\end{equation}
realized through the action \eqref{efff} of \cite{Sfetsos:2013wia}
\begin{equation}
e=2d(\lambda)=\frac{1}{\sqrt{k(1-\lambda^2)}}\,\frac{2\lambda}{1+\lambda}\,,\quad
\frac{c(\lambda)}{d(\lambda)}=1+2x\,,\quad x=\frac{1+\lambda^2}{2\lambda}\,.
\end{equation}
That the deformed brackets \eqn{defpoira} follow as the classical limit of the
OPEs provides actually, for the isotropic case,
the mathematical proof that the action \eqn{efff} is in fact the effective action of the non-Abelian
Thirring model action \eqn{WZW-pert}. The reason is that \eqn{efff} provides,
as was shown in \cite{Sfetsos:2013wia},
a realization of \eqn{defpoira} which in turn was derived
by using \eqn{WZW-pert} as the starting point.

\section{Conclusions}
\label{conclusion}

In this work we have computed all possible two- and three-point functions of current
and primary field operators for the $\lambda$-deformed integrable $\sigma$-models.
These models are characterised by the deformation parameter $\lambda$,
as well as by the integer level $k$ of the WZW model.
Our results are valid for any semisimple group $G$, for all values of the
deformation parameter $\lambda$ and up to order $1/k$ in the large $k$ expansion.
We achieved this goal by combining the first few orders in perturbation theory
with analyticity arguments as well as
with a non-trivial duality-type symmetry shared by these models.
The two- and three-point correlators
allowed us to deduce the exact in $\lambda$ OPEs of all currents and primary operators.
Furthermore, based on our results we derived the anomalous dimensions and correlation functions
for the operators in two important limits of the aforementioned $\lambda$-deformed $\sigma$-models, namely
the non-Abelian T-dual of the PCM and the pseudodual model.

\no
Our results are summarized as follows:

\begin{enumerate}

\item
In section \ref{sec:jj2} we presented the results for the two-point correlator of two currents.
From these correlators and in conjunction with
the aforementioned symmetry we derived the all-loop $\b$-function of the theory as well as their anomalous dimension.

\item
In section \ref{sec:jjj3} we derived the all-loop expressions for the three current correlators.

\item
In section \ref{sec:prim2} we provide the reader with the exact two-point functions of all primary operators of the theory, as well as with
their exact in $\lambda$ anomalous dimensions. In this case, the role of the symmetry is instrumental since it is realised in a non-trivial way.

\item
In section \ref{sec:prim3} we provide the reader with the exact three-point functions of all primary operators of the theory.

\item
In section \ref{sec:mixed} we calculated the exact, in $\lambda$, three-point correlators
$\langle J \Phi \Phi\rangle$ and $\langle \bar J \Phi \Phi\rangle$.

\item
In section \ref{sec:OPE} we deduced all relevant OPEs between
currents and/or primary fields that are consistent with the
exact results for the two- and three-point functions given in
previous sections.
We also derive the currents' Poisson brackets which assume Rajeev's {\it deformation of the
canonical structure of the isotropic PCM}, the underlying structure of the integrable $\lambda$-deformed
$\sigma$-models. This essentially proves in a mathematical sense
that the action \eqn{efff} for an isotropic deformation is indeed the effective action of the
non-Abelian Thirring model action.

\end{enumerate}

One direction for extending our work would be to consider cases beyond isotropy,
i.e. when the matrix $\lambda$ is not proportional to the identity. 
In particular, we believe that the equal-time commutators of the currents and primaries will take the form of \eqref{ssss},
under an analogue to \eqref{jkfllfld} relation.
Another direction would be to calculate the subleading, in the $1/k$ expansion,
terms of all physical quantities such as the $\b$-function,
the anomalous dimension matrix and the fusion coefficients.
These line of research  would, hopefully, be culminated by finding the exact
in both $\lambda$ and $k$ expressions
for these physical quantities as well as the underlying effective action.

\section*{Acknowledgments}

The authors would like to thank each others home institutes for hospitality.
The research of K. Siampos is partially supported by the \textsl{Germaine de Stael}
France--Swiss bilateral program (project no 32753SG).
K. Sfetsos and K. Siampos would like to thank the TH-Unit at CERN for hospitality and financial support
during the final stages of this project.
K. Siampos would like also to thank the ICTP, Trieste for hospitality during the final stages of this project.

\begin{appendices}

\section{Various integrals}
\label{vint}

In this appendix we assemble all the integrals that will be needed in our perturbative calculations.
In all integrals considered below the integration domain is a disc of radius $R$ in which the various external
points labeled by $x$'s are excluded. This can be done by encircling them with circles having
arbitrarily vanishing radius.
One way to prove the expressions below is to use Stokes' theorem in two-dimensions for
appropriately chosen vectors and contours.

\no
The first set of integrals is the exact version of the integrals in \eqn{id1} and \eqn{id3}
\begin{equation}
\label{Kutasov}
\begin{split}
&\int\frac{\text{d}^2z}{(z-x_1)(\bar z-\bar x_2)}=-\pi\,\ln\frac{|x_1-x_2|^2}{R^2-x_1\bar x_2}\ ,
\\
&\int\frac{\text{d}^2z}{(z-x_1)^2(\bar z-\bar x_2)^2}=\pi^2\delta^{(2)}(x_1-x_2)-\frac{\pi R^2}{(R^2-x_1\bar x_2)^2}\, ,
\end{split}
\end{equation}
where the $R>|x_{1,2}|$.
By taking derivatives we may compute the exact analog of the integrals in \eqn{id2}.

\no
A generalization of the first of the above integrals is given by
\begin{equation}
\int\frac{\text{d}^2z}{\prod\limits_{i=1}^M(z-x_i)\,\prod\limits_{i=1}^N(\bar z-\bar y_j)}=
-\pi\sum\limits_{i=1}^M\sum\limits_{j=1}^N\,\frac{1}{A_i B_j}\,\ln\frac{|x_i-y_j|^2}{R^2-x_i\bar y_j}\,,
\end{equation}
with $R>\{|x_i|,|y_j|\}$ and $A_i=\prod\limits_{\substack{j=1 \\ j\neq i}}^M(x_i-x_j)\,,\quad
B_i=\prod\limits_{\substack{j=1 \\ j\neq i}}^N(\bar y_i-\bar y_j)$.
This relation can be proved by first performing
a partial fraction decomposition and then use \eqn{Kutasov}.
A special case of this is when the denominators are cubic polynomials. Namely,
\begin{equation}
\begin{split}
&
\int {\text{d}^2z\ov (z-x_1)(z-x_2)(\bar z-\bar x_1)}=  -{\pi \ov x_{12}} \left(\ln {\varepsilon^2\ov |x_{12}|^2} + \ln {R^2-x_2 \bar x_1\ov R^2 - |x_1|^2}\right)\ ,
\\
&
\int {\text{d}^2z\ov (z-x_1)(\bar z-x_1)(\bar z-\bar x_2)} = -{\pi \ov \bar x_{12}}
 \left(\ln {\varepsilon^2\ov |x_{12}|^2} + \ln {R^2-x_1 \bar x_2\ov R^2 - |x_1|^2}\right)\ ,
\end{split}
\end{equation}
which is the exact analogs of the integrals in \eqn{id22}.

\no
Another important  integral is given by
\begin{equation}
\int\frac{\text{d}^2z}{(z-x_1)(\bar z-\bar x_2)}\,\ln\frac{|z-x_1|^2}{R^2-x_1\bar z}=
-\frac\pi2\ln^2\frac{|x_1-x_2|^2}{R^2-x_1\bar x_2}\ .
\label{logp}
\end{equation}
The large $R$ limit of this is given in \eqn{logpla} and is necessary
for the derivation of the two-loop contribution in \eqref{JbJaiso}.

\section{Perturbative computation of the $\langle J\bar J\rangle$ correlator}
\label{detaJbarJ}

In this appendix we present the perturbative calculation of the $\langle J\bar J\rangle$ two-point function.
At the conformal point it vanishes.

\no
{\bf Order ${\cal O}(\l)$:} To that order we have that
\begin{equation}
\langle J^a(x_1)J^b(x_2)\rangle_\l^{(1)} = -{\l \ov \pi} \int \text{d}^2z \langle J^a(x_1) J^c(z)\rangle \langle \bar J^c(\bar z) \bar J^b(\bar x_2)\rangle =
-\pi \l \d^{ab} \d^{(2)}(x_{12})\ .
\label{jjex1}
\end{equation}

\no
{\bf Order ${\cal O}(\l^2)$:} To that order we have that
\ba
&& \langle J^a(x_1)J^b(x_2)\rangle_\l^{(2)} = {\l^2 \ov2  \pi^2} \int \text{d}^2z_{12}
\langle J^a(x_1) J^{a_1}(z_1) J^{a_2}(z_2)\rangle \langle \bar J^{a_1}(\bar z_1) \bar J^{a_2}(\bar z_2) \bar J^b(\bar x_2)\rangle
\nonumber\\
&&\qq\qq \phantom{xxxxx} = - {\l^2\ov 2\pi^2} \d^{ab} {c_G\ov k} J(x_1,x_2)\ ,
\label{jjex2}
\ea
where
\ba
&& J(x_1,x_2) =  \int {\text{d}^2z_{12}\ov (x_1-z_1) (x_1-z_2) (z_1-z_2) (\bar x_2-\bar z_1)
(\bar x_2-\bar z_2) (\bar z_1-\bar z_2)}
\nonumber \\
&& \quad =  \int {\text{d}^2z_{12}\ov (x_1-z_1)^2 (\bar x_2-\bar z_2)^2 }
\left( {1\ov z_1-z_2}-{1\ov x_1-z_2}\right) \left( {1\ov \bar z_1-\bar z_2}+{1\ov \bar x_2-\bar z_1}\right)
\\
&&  \quad = J_{11} + J_{12} + J_{21} + J_{22}\ ,
\nonumber\ea
where we have broken the integral $J$ into the four integrals $J_{ij}$, resulting from
multiplying out the terms in the parenthesis, in a rather self-explanatory notation.
We have that
\ba
&& J_{11} = \del_{x_1}\del_{\bar x_2} \int {\text{d}^2z_{12}\ov (x_1-z_1) (\bar x_2-\bar z_2) (z_1-z_2)(\bar z_1 - \bar z_2)}
\nonumber
 \label{j11}
 \\
&&\quad   =  -\pi  \del_{x_1}\del_{\bar x_2} \int {\text{d}^2 z_2\ov (x_1-z_2)(\bar x_2-\bar z_2)} \ln {\varepsilon^2\ov |z_2-x_1|^2}
\nonumber\\
&&\quad  = \pi^2 \del_{x_1}\del_{\bar x_2}  \left(  \ln\varepsilon^2\ln|x_{12}|^2  - \ha \ln^2|x_{12}|^2\right)
\\
&& \quad  = -\pi^3 \d^{(2)}(x_{12}) \ln {\varepsilon^2\ov |x_{12}|^2} +{\pi^2\ov |x_{12}|^2}\ ,
\nonumber
\ea
where we have used \eqn{logpla}. Also
\begin{equation}
\begin{split}
& J_{12} = \pi \int {\text{d}^2 z_1\ov (x_1-z_1)^2 (\bar x_2-\bar z_1)^2} = \pi^3 \d^{(2)}(x_{12})\ ,
\label{j1221}
\\
&J_{21} = \pi \int {\text{d}^2z_2\ov (\bar x_2-\bar z_2)^2 (x_1-z_2)^2} = \pi^3 \d^{(2)}(x_{12})\ .
\end{split}
\end{equation}
Note that for $J_{12}$ we have first performed the $z_2$-integration which is not in accordance with 
our regularization prescription. However, we now show that the same result follows if we do first the $z_1$
according to our regularization. We easily find that
\ba
&& J_{12}= \del_{x_1} \int {\text{d}^2z_{12}\ov (x_1-z_2)(\bar x_2 - \bar z_2)^2}
\left({1\ov (z_1-x_1)(\bar x_2-\bar z_1)} - {1\ov (z_1-z_2)(\bar x_2-\bar z_2)} \right)
\nonumber \\
&&= \pi \del_{x_1} \left(\ln |x_{12}|^2 \int {\text{d}^2z_2\ov (\bar x_2-\bar z_2)^2(x_1-z_2)}\right)
+ \pi \int {\text{d}^2z_2 \ln|z_2-x_2|^2 \ov (z_2-x_1)^2 (\bar z_2- \bar x_2)^2}
\\
&& = -\pi^2 \del_{x_1}\left( \ln |x_{12}|^2\ov \bar x_{12}\right) + \pi^3 (1+\ln|x_{12}|^2 )\,\d^{(2)}(x_{12}) + {\pi^2\ov |x_{12}|^2}\ ,
\nonumber
\ea
where we have used the fact that the second integral in the second line above can be obtained from \eqn{j11} (with $\varepsilon=1$).
A simple algebra gives the same expression as in \eqn{j1221}.
Finally
\begin{equation}
J_{22} =-{\pi\ov x_{12}} \int {\text{d}^2z_2\ov  (\bar x_2-\bar z_2)^2 ( x_1 - z_2)} = {\pi^2\ov |x_{12}|^2}\ .
\end{equation}
Therefore collecting all contributions we find that
\begin{equation}
J(x_1,x_2) = 2\pi^4\left(1-\ha  \ln {\varepsilon^2\ov |x_{12}|^2}\right) \d^{(2)}(x_{12}) + {\pi^3\ov |x_{12}|^2}\ .
\end{equation}

\no
{\bf Order ${\cal O}(\l^3)$:} To that order we have that
\begin{equation}
\begin{split}
& \langle J^a(x_1)J^b(x_2)\rangle_\l^{(3)} = - {\l^3 \ov 6  \pi^3} \int \text{d}^2z_{123}
\langle J^a(x_1) J^{a_1}(z_1) J^{a_2}(z_2) J^{a_3}(z_3)\rangle\times
\\
&\hskip 5.5 cm
 \langle \bar J^{a_1}(\bar z_1) \bar J^{a_2}(\bar z_2) \bar J^{a_3}(\bar z_3) \bar J^b(\bar x_2)\rangle
\end{split}
\end{equation}
The four-point function is given by \eqn{jp4} and is
multiplied by the analogous four-point function for antiholomorphic currents.
Keeping terms up to ${\cal O}(1/k)$, disregarding terms giving rise to bubbles and taking into account the above permutation symmetry
we arrive at
\begin{equation}
\langle J^a(x_1)J^b(x_2)\rangle_\l^{(3)} =  2\d^{ab}\, {c_G\ov k}\,{\l^3 \ov   \pi^3} K(x_1,x_2)\ ,
\label{jjex3}
\end{equation}
where
\begin{equation}
K(x_1,x_2)=
\int {\text{d}^2z_{123} \ov (z_1-x_2)(x_1-z_2)(x_1-z_3)(z_2-z_3) (\bar z_1-\bar z_2)^2 (\bar z_3 -\bar x_2)^2}\ .
\end{equation}
Performing the integrations first over $z_1$ and then over $z_2$ we obtain that
\ba
&& K(x_1,x_2)= -\pi^2 \int {\text{d}^2z_3 \ov (z_3 - x_1)^2 (\bar z_3-\bar x_2)^2} \ln {\varepsilon^2\ov |z_3-x_1|^2}
= -\pi^4 \ln \varepsilon^2\d^{(2)}(x_{12})
\nonumber\\
&&  + \pi^2\left( \del_{x_1} \del_{\bar x_2} \int {\text{d}^2z_3 \ov (z_3 - x_1) (\bar z_3-\bar x_2)} \ln |z_3-x_1|^2 +
 \int {\text{d}^2z_3\ov (z_3-x_1)^2(\bar z_3-\bar x_2)^2} \right)
\nonumber\\
&& = -\pi^4 \ln \varepsilon^2\d^{(2)}(x_{12}) + \pi^4 \d^{(2)}(x_{12}) -{\pi^3\ov 2} \del_{x_1} \del_{\bar x_2} \ln^2 |x_{12}|^2
\\
&& = \pi^4\left(1- \ln {\varepsilon^2\ov |x_{12}|^2}\right) \d^{(2)}(x_{12}) + {\pi^3\ov |x_{12}|^2}\ .
\nonumber
\ea
In conclusion \eqn{jjex1}, \eqn{jjex2} and \eqn{jjex3} combine to
\eqn{JbJaiso} in the main text.

\section{Perturbative computation of the $\langle JJJ\rangle$ correlator}
\label{jjjcor}

In this appendix we present the perturbative calculation of the  $\langle JJJ\rangle$ three-point correlator.
The ${\cal O}(\l)$ contribution to this correlators vanishes since $\langle \bar J\rangle=0$.
Proceeding to higher orders in the $\lambda$-expansion
we have:

\no
{\bf Order ${\cal O}(\l^2)$:} This contribution is immediately seen to be equal to
\begin{equation}
\label{jfkdk1}
\langle J^a(x_1)J^b(x_2)J^c(x_3)\rangle^{(2)}_\l
=\frac{\lambda^2}{2!\pi^2}\int {\text{d}^2z_{12}\ov \bar z^2_{12}}
\langle J^a(x_1)J^b(x_2)J^c(x_3)J^{a_1}(z_1)J^{a_1}(z_2)\rangle\ .
\end{equation}
To proceed with the contractions we single out $J^{a_1}$ to perform them. Disregarding the disconnected and bubble pieces
and also
noting that the Abelian contractions, i.e. contractions leading to second poles, of $J^{a_1}$ with the external currents vanish
in our regularization scheme,
we have that
\ba
\label{jjjex2}
&& \langle J^a(x_1)J^b(x_2)J^c(x_3)\rangle^{(2)}_\l=\frac{\lambda^2}{2!\pi^2}\frac{1}{\sqrt{k}}\int\frac{\text{d}^2z_{12}}{\bar z_{12}^2}
\left(\frac{f_{a_1ad}}{z_1-x_1}\langle J_d(x_1)J_b(x_2)J_c(x_3)J_{a_1}(z_2)\rangle\right.
\nonumber
\\
&&\left.+\frac{f_{a_1bd}}{z_1-x_2}\langle J_a(x_1)J_d(x_2)J_c(x_3)J_{a_1}(z_2)\rangle
+\frac{f_{a_1cd}}{z_1-x_3}\langle J_a(x_1)J_b(x_2)J_d(x_3)J_{a_1}(z_2)\rangle
\right)
\nonumber \\
&&=\frac{\lambda^2}{2\pi^2}\frac{1}{\sqrt{k}}\int\text{d}^2z_{2}
\left(\frac{f_{a_1ad}}{\bar x_1-\bar z_2}\langle J_d(x_1)J_b(x_2)J_c(x_3)J_{a_1}(z_2)\rangle\right.
\\
&&\left.+\frac{f_{a_1bd}}{\bar x_2-\bar z_2}\langle J_a(x_1)J_d(x_2)J_c(x_3)J_{a_1}(z_2)\rangle
+\frac{f_{a_1cd}}{\bar x_3-\bar z_2}\langle J_a(x_1)J_b(x_2)J_d(x_3)J_{a_1}(z_2)\rangle
\right)\,.
\nonumber
\ea
Computing this to ${\cal O}\left(1/\sqrt{k}\right)$ gives
\begin{equation}
\langle J^a(x_1)J^b(x_2)J^c(x_3)\rangle^{(2)} =\frac{3\lambda^2}{2\sqrt{k}}\,\frac{f_{abc}}{x_{12}x_{13}x_{23}}\,.
\end{equation}

\no
{\bf Order ${\cal O}(\l^3)$:} The contribution is immediately seen to be
\ba
\label{jfkdk2}
&& \langle J^a(x_1)J^b(x_2)J^c(x_3)\rangle^{(3)}=-\frac{\lambda^3}{3!\pi^3}\frac{f_{a_1a_2a_3}}{\sqrt{k}}\int {\text{d}^2z_{123} \ov
\bar z_{12}\bar z_{13}\bar z_{23}}\times
\nonumber\\
&&\qq\qq\qq\qq  \langle J_a(x_1)J_b(x_2)J_c(x_3)J_{a_1}(z_1)J_{a_2}(z_2)J_{a_3}(z_3)\rangle\ .
\ea
As we have already saturated the ${\cal O}(1/\sqrt{k})$,
we perform only Abelian contractions in this six-point function, yielding to
\begin{equation}
\begin{split}
\langle J^a(x_1)J^b(x_2)J^c(x_3)\rangle^{(3)}_\l=
-\frac{\lambda^3f_{abc}}{\pi^3\sqrt{k}}\int
\frac{ \text{d}^2z_{123}
  }{\bar z_{12}\bar z_{13}\bar z_{23}(z_1-x_1)^2(z_2-x_2)^2(z_3-x_3)^2}\,.
\end{split}
\end{equation}
Using the identity
\begin{equation}
\label{jfkkf}
\frac{1}{\bar z_{12}\bar z_{13}}=\frac{1}{\bar z_{23}}\left(\frac{1}{\bar z_{12}}-\frac{1}{\bar z_{13}}\right)\, ,
\end{equation}
integrating over $z_1$
\begin{equation}
\begin{split}
\langle J^a(x_1)J^b(x_2)J^c(x_3)\rangle^{(3)}_\l=
-\frac{\lambda^3f_{abc}}{\pi^2\sqrt{k}}\partial_{x_2}\partial_{x_3}\int
\frac{\text{d}^2z_{23}}{\bar z^2_{23}(z_2-x_2)(z_3-x_3)}\left(\frac{1}{z_2-x_1}-
\frac{1}{z_3-x_1}\right)\
\end{split}
\end{equation}
and employing an analogue of the identity \eqref{jfkkf} we get that
\begin{equation}
\begin{split}
& \langle J^a(x_1)J^b(x_2)J^c(x_3)\rangle^{(3)}_\l=
-\frac{\lambda^3f_{abc}}{\sqrt{k}}\partial_{x_2}\partial_{x_3}
\left(\frac{1}{x_{12}}\ln\frac{|x_{13}|^2}{|x_{23}|^2}-\frac{1}{x_{13}}\ln\frac{|x_{12}|^2}{|x_{23}|^2}\right)
\\
&\qq\qq\qq\qq \qq = -\frac{\lambda^3}{\sqrt{k}}\,\frac{f_{abc}}{x_{12}x_{13}x_{23}}\ .
\end{split}
\end{equation}

\section{Perturbative computation of the $\langle JJ\bar J\rangle$ correlator}
\label{jjbjcor}

 In this appendix we present the perturbative calculation of the  $\langle JJ\bar J\rangle$ three-point correlator.
Of course at the conformal point this correlation function vanishes.
Proceeding to higher orders in the $\lambda$-expansions we have that:

\no
{\bf Order ${\cal O}(\l)$:}
The contribution to the one-loop equals
\begin{equation}
\label{jfkdk3}
\begin{split}
\langle J^a(x_1)J^b(x_2) \bar J^c(\bar x_3)\rangle^{(1)}_\l
&=-\frac{\lambda}{\pi}\int\text{d}^2z
\langle J^a(x_1)J^b(x_2)J^{a_1}(z)\rangle
\langle \bar J^{a_1}(\bar z)\bar J^{c}(\bar x_3)\rangle\,,\\
&=-\frac{\lambda f_{abc}}{\sqrt{k}\,x_{12}}\int\frac{\text{d}^2z}{(x_1-z)(x_2-z)(\bar x_3-\bar z)^2}\, .
\end{split}
\end{equation}
Employing an analogue of the identity \eqref{jfkkf} we get that
\begin{equation}
\langle J^a(x_1)J^b(x_2) \bar J^c(\bar x_3)\rangle^{(1)}_\l
=\frac{\lambda}{\sqrt{k}}\,\frac{f_{abc} \bar x_{12}}{x_{12}^2\bar x_{23}\bar x_{13}}\,.
\end{equation}

\no
{\bf Order ${\cal O}(\l^2)$:}
The contribution to the two-loop is equal to
\begin{equation}
\begin{split}
\langle J^a(x_1)J^b(x_2) \bar J^c(\bar x_3)\rangle^{(2)}_{abc}
&=\frac{\lambda^2f_{a_1a_2c}}{2!\pi^2\sqrt{k}}\int\text{d}^2z_{12}
\frac{\langle J_a(x_1)J_b(x_2)J_{a_1}(z_1)J_{a_2}(z_2)\rangle}{\bar z_{12}(\bar z_1-\bar x_3)(\bar z_2-\bar x_3)}\\
&=\frac{\lambda^2f_{abc}}{\pi^2\sqrt{k}}
\int\frac{\text{d}^2z_{12}}{\bar z_{12}(\bar z_1-\bar x_3)(\bar z_2-\bar x_3)(x_1-z_1)^2(x_2-z_2)^2}\ .
\end{split}
\end{equation}
Employing again an analogue of the identity \eqref{jfkkf} we get that
\ba
\langle J^a(x_1)J^b(x_2) \bar J^c(\bar x_3)\rangle^{(2)}_{abc}
&& = \frac{\lambda^2f_{abc}}{\pi^2\sqrt{k}}
\int\frac{\text{d}^2z_{12}}{(\bar z_2-\bar x_3)^2(x_1-z_1)^2(x_2-z_2)^2}
\left(\frac{1}{\bar z_1-\bar z_2}-\frac{1}{\bar z_1-\bar x_3}\right)
\nonumber
\\
&& =-\frac{\lambda^2f_{abc}}{\pi\sqrt{k}}
\int\frac{\text{d}^2z_{2}}{(\bar z_2-\bar x_3)^2(x_2-z_2)^2}
\left(\frac{1}{x_1- z_2}-\frac{1}{x_{13}}\right)
\nonumber
\\
&& =\frac{\lambda^2f_{abc}}{\pi\sqrt{k}}\partial_{x_2}\partial_{\bar x_3}
\int\frac{\text{d}^2z_{2}}{(x_1-z_2)(\bar z_2-\bar x_3)(x_2-z_2)}
\\
&& =\frac{\lambda^2f_{abc}}{\pi\sqrt{k}}\partial_{x_2}\partial_{\bar x_3}
\frac{1}{x_{12}}\int\frac{\text{d}^2z_{2}}{\bar z_2-\bar x_3}
\left(\frac{1}{x_2-z_2}-\frac{1}{x_1-z_2}\right)
\nonumber
\\
&& =\frac{\lambda^2f_{abc}}{\sqrt{k}}
\partial_{x_2}\partial_{\bar x_3}\frac{1}{x_{12}}\ln\frac{|x_{23}|^2}{|x_{13}|^2}\\
&&= -\frac{\lambda^2}{\sqrt{k}} f_{abc}
\left(\frac{\bar x_{12}}{x_{12}^2\bar x_{23}\bar x_{13}}
+\pi {\d^{(2)}(x_{23})
\ov x_{12}}\right)\ ,
\nonumber
\ea
where we have included the contact term involving external points.
This will be neglected in the main text.

\section{Perturbative computation of the $\langle\Phi \Phi\rangle$ correlator}
\label{ffcor}

In this appendix we present the perturbative calculation of the $\langle\Phi \Phi\rangle$ correlator.\\
{\bf Order ${\cal O}(\l)$:} To that order we have that
\ba
&& \langle \Phi^{(1)}_{i,i'}(x_1,\bar x_1) \Phi^{(2)}_{j,j'}(x_2, \bar x_2)\rangle_\l^{(1)}
= -{\l \ov \pi} \int \text{d}^2z \langle \Phi^{(1)}_{i,i'}(x_1,\bar x_1) J^c(z) \bar J^c(\bar z)
\Phi^{(2)}_{j,j'}(x_2,\bar x_2)\rangle =
\nonumber\\
&&\phantom{xxxxxxxxx} = {\l\ov \pi \sqrt{k}} \int \text{d}^2 z \Big[ {(t_a^{(1)})_i{}^\ell \ov z-x_1}
\langle \Phi^{(1)}_{\ell,i'}(x_1,\bar x_1) \bar J^c(\bar z) \Phi^{(2)}_{j,j'}(x_2,\bar x_2)\rangle
\nonumber\\
&& \phantom{xxxxxxxxxxx}  + {(t_a^{(2)})_j{}^\ell \ov z-x_2}
\langle \Phi^{(1)}_{i,i'}(x_1,\bar x_1) \bar J^c(\bar z) \Phi^{(2)}_{\ell,j'}(x_2,\bar x_2)\rangle\Big]
\\
&& = -{\l/k\ov x_{12}^{2 \D_R} \bar x_{12}^{2 \bar \D_{R'}}}
\Big[\ln \varepsilon^2\left(t_a^{(1)} \otimes \tilde t_a^{(1) T} + t_a^{(2)T} \otimes \tilde t_a^{(2)}\right)
\nonumber
\\
&& \qq\qq\qq
+ \ln |x_{12}|^2 \left(t_a^{(1)} \otimes \tilde t_a^{(2)} + t_a^{(1)T} \otimes \tilde t_a^{(2)T}\right)\Big]_{ii',jj'}\ ,
\nonumber
\ea
where we have used \eqn{JJf1} and \eqn{JJf2} and wrote the result as a direct product of matrices.
The representations involved are in fact conjugate to each other. Therefore, using \eqn{conjj} we find that
\begin{equation}
\langle \Phi^{(1)}_{i,i'}(x_1,\bar x_1) \Phi^{(2)}_{j,j'}(x_2, \bar x_2)\rangle_\l^{(1)}
= -2 {\l\ov  k} {(t_a \otimes t^*_a)_{ii',jj'}\ov x_{12}^{2 \D_R} \bar x_{12}^{2 \bar \D_{R'}}}
\ln {\varepsilon^2\ov |x_{12}|^2}\ .
\end{equation}

\no
{\bf Order ${\cal O}(\l^2)$:} To that order we find that
\begin{equation}
\langle \Phi^{(1)}_{i,i'}(x_1,\bar x_1) \Phi^{(2)}_{j,j'}(x_2, \bar x_2)\rangle_\l^{(2)}= {\l^2\ov 2\pi^2}
\left[{1\ov \sqrt{k}}\left(I^{(1)}_{ii',jj'}+ I^{(2)}_{ii',jj'} + I^{(3)}_{ii',jj'}\right) + I^{(4)}_{ii',jj'}\right]\ ,
\end{equation}
where the four different terms are computed below and arise by using the current Ward identity
with respect to the current $J_a(z_1)$.

\no
$\bullet$ The first term is
\ba
&& I^{(1)}_{ii',jj'} = -\int \text{d}^2z_{12}{(t^{(1)}_a)_i{}^\ell\ov z_1-x_1} \langle \Phi^{(1)}_{\ell,i'}(x_1.\bar x_1)
\bar J_a(\bar z_1) J_b(z_2) \bar J_b(\bar z_2) \Phi^{(2)}_{j,j'}(x_2.\bar x_2)\rangle
\nonumber\\
&& = -\int \text{d}^2z_{12}{(t^{(1)}_a)_i{}^\ell\ov (z_1-x_1) \bar z_{12}^2}
\langle \Phi^{(1)}_{\ell,i'}(x_1,\bar x_1) J_a(z_2) \Phi^{(2)}_{j,j'}(x_2,\bar x_2)\rangle
\\
&& =\pi^2 {C_R\ov \sqrt{k}} {(\mathbb{I}_R \otimes \mathbb{I}_{R'})_{ii',jj'} \ov x_{12}^{2 \D_R} \bar x_{12}^{2 \bar \D_{R'}}}
\ln {\varepsilon^2\ov |x_{12}|^2}\ ,
\nonumber
\ea
where we have kept only contributions which will give terms of ${\cal O}(1/k)$ to the final result.
In addition we used the integral
\begin{equation}
\int {\text{d}^2z_{12}\ov (z_1-x_1)(z_2-x_2)\bar z_{12}^2}
=\pi  \int {\text{d}^2z_2\ov (\bar x_1-\bar z_2)(z_2-x_2)}=\pi^2\ln |x_{12}|^2\ ,
\end{equation}
as well as the same with $x_2\to x_1$ in which case $|x_{12}|^2\to \varepsilon^2$ in the result above.

\no
$\bullet$ The second term is
\ba
I^{(2)}_{ii',jj'} &&= -\int \text{d}^2z_{12}{(t^{(2)}_a)_j{}^\ell\ov z_1-x_2} \langle \Phi^{(1)}_{i,i'}(x_1,\bar x_1)
\bar J_a(\bar z_1) J_b(z_2) \bar J_b(\bar z_2) \Phi^{(2)}_{\ell,j'}(x_2,\bar x_2)\rangle
\nonumber\\
&& = -\int \text{d}^2z_{12}{(t^{(2)}_a)_j{}^\ell\ov (z_1-x_2) \bar z_{12}^2}
\langle \Phi^{(1)}_{i,i'}(x_1,\bar x_1) J_a(z_2) \Phi^{(2)}_{\ell,j'}(x_2,\bar x_2)\rangle
\\
&& =\pi^2 {C_R\ov \sqrt{k}} {(\mathbb{I}_R \otimes \mathbb{I}_{R'})_{ii',jj'} \ov x_{12}^{2 \D_R} \bar x_{12}^{2 \bar \D_{R'}}}
\ln {\varepsilon^2\ov |x_{12}|^2}\ ,
\nonumber
\ea
where as before we have kept only contributions providing at most ${\cal O}(1/k)$ terms in the final result.

\no
$\bullet$ The third term is
\ba
I^{(3)}_{ii',jj'} = f_{abc} \int {\text{d}^2z_{12}\ov z_{12}} \langle \Phi^{(1)}_{i,i'}(x_1,\bar x_1)
\bar J_a(\bar z_1) J_c(z_2) \bar J_b(\bar z_2) \Phi^{(2)}_{\ell,j'}(x_2,\bar x_2)\rangle=0\,,
\ea
since to ${\cal O}(1/\sqrt{k})$ we get a result proportional to $f_{abc}\,\delta_{ab} = 0 $.

\no
$\bullet$ The fourth term is more involved to compute. The result is
\ba
&& I^{(4)}_{ii',jj'} =  \int {\text{d}^2z_{12}\ov z_{12}^2}
\langle \Phi^{(1)}_{i,i'}(x_1,\bar x_1)
\bar J_a(\bar z_1) \bar J_a(\bar z_2) \Phi^{(2)}_{\ell,j'}(x_2,\bar x_2)\rangle
\nonumber\\
&& \qq\ = 2 \pi^2 {C_{R'}\ov k} {(\mathbb{I}_R \otimes \mathbb{I}_{R'})_{ii',jj'} \ov x_{12}^{2 \D_R} \bar x_{12}^{2 \bar \D_{R'}}}
\ln {\varepsilon^2\ov |x_{12}|^2}\ .
\ea
Note that this is expected since it is just the sum of the other two non-vanishing terms with the representations
$R$ and $R'$ exchanged.

\no
{\bf Order ${\cal O}(\l^3)$:} To that order we have that
\begin{equation}
\langle \Phi^{(1)}_{i,i'}(x_1,\bar x_1) \Phi^{(2)}_{j,j'}(x_2, \bar x_2)\rangle_\l^{(3)}= -{\l^3\ov 6\pi^3}
\left[J^{(1)}_{ii',jj'}+ J^{(2)}_{ii',jj'} + J^{(3)}_{ii',jj'} + J^{(4)}_{ii',jj'}+J^{(5)}_{ii',jj'}+J^{(6)}_{ii',jj'}\right]\ ,
\end{equation}
where the six different terms are obtained by applying the Ward identity for the current $J_a(z_1)$.

\no
$\bullet$ The first term originates from the contraction of the current $J_a(z_1)$ with the primary field $\Phi^{(1)}$
and leads to
\ba
&& J^{(1)}_{ii',jj'} = -{1\ov \sqrt{k}}\int \text{d}^2z_{123} {(t^{(1)}_a)_i{}^k\ov z_1-x_1}
\nonumber\\
&&\qq \quad  \langle \Phi^{(1)}_{k,i'}(x_1,\bar x_1)
\bar J_a(\bar z_1) J_b(z_2) \bar J_b(\bar z_2) J_c(z_3) \bar J_c(\bar z_3) \Phi^{(2)}_{j,j'}(x_2,\bar x_2)\rangle
\ea
The next step is to contract one of the remaining holomorphic currents, lets say $J_b(z_2)$.
This current can not be contracted with any of the external primaries because in that
case the last holomorphic current should also be contracted with an external field too and as a result this contribution will
be of order $1/k^{3/2}$. Since in this calculation we keep terms of order ${\cal O}(1/k)$ this contribution can be ignored.
For the same reason  the holomorphic currents $J_b(z_2)$ and $J_c(z_3)$ can not be contracted through the non-Abelian part
of their OPE but only via the Abelian part.
Once we have contracted all the holomorphic currents we start treating the anti-holomorphic ones
by choosing $\bar J_a(\bar z_1)$ to use in the Ward identity. As above, this current cannot be contracted with any of the external primaries
since in this case the remaining anti-holomorphic currents at $\bar z_2$ and at $\bar z_3$ should be contracted through a {$\delta$-Kronecker term} and
resulting into the term ${1 \ov z_{23}^2\bar z_{23}^2}$ which indicates that it is disconnected and should be ignored.
Thus, the current at $\bar z_1$ can be contracted only with the anti-holomorphic currents at $\bar z_2$ and at $\bar z_3$.
Notice, however, that this contraction can not be non-Abelian because in that case the result will be proportional
to $f_{abc} \delta_{bc}=0$.
We have thus concluded that the only non vanishing terms up to order ${\cal O}(1/k)$ will come from the Abelian contractions of $\bar J_a(\bar z_1)$
with the other anti-holomorphic currents. The resulting integral is
\ba
\label{J1}
&& J^{(1)}_{ii',jj'} = -{2\ov k}{1 \ov x_{12}^{2\D_R} \bar x_{12}^{2\D_{R'}}}\int \text{d}^2z_{123}
\Big( {\big( t_a^{(1)} \otimes \tilde t_a^{(1) T} \big)_{ii',jj'}  \ov z_{23}^2 \bar z_{12}^2 (z_1-x_1)(\bar z_3-\bar x_1)}
\nonumber \\
&& \qq + {\big( t_a^{(1)} \otimes \tilde t_a^{(2) } \big)_{ii',jj'}  \ov z_{23}^2
\bar z_{12}^2 (z_1-x_1)(\bar z_3-\bar x_2)}+ (z_2 \leftrightarrow z_3) \Big)\ ,
\ea
where the $z_2,z_3$ exchange term applies only in the integrand and not in the measure of integration in accordance with our regularization prescription.
It turns out that this term doubles the result of the term explicitly written. The integrals can now be performed from the left to the right, the $z_1$ integration first then $z_2$  and the $z_3$ last. Using \eqref{conjj} the result can be written as follows
\be
J^{(1)}_{ii',jj'} =2 {\pi^3\ov  k} {(t_a \otimes t^*_a)_{ii',jj'}\ov x_{12}^{2 \D_R} \bar x_{12}^{2 \bar \D_{R'}}}
\ln {\varepsilon^2\ov |x_{12}|^2}\ .
\ee
We should mention that the $\ln {\varepsilon^2}$ term originates from the first triple integral  of \eqref{J1} while the $\ln {|x_{12}|^2}$ term originates
the second triple integral  of \eqref{J1}.

\no
$\bullet$ The second term originates from the contraction of the current $J_a(z_1)$ with $J_b(z_2)$
through the non-Abelian term of their OPE.
It reads
\be
J^{(2)}_{ii',jj'} = {f_{abd}\ov \sqrt{k}}\int {\text{d}^2z_{123} \ov z_{12}} \langle \Phi^{(1)}_{i,i'}(x_1,\bar x_1)
\bar J_a(\bar z_1) J_d(z_2) \bar J_b(\bar z_2) J_c(z_3) \bar J_c(\bar z_3) \Phi^{(2)}_{j,j'}(x_2,\bar x_2)\rangle \ .
\ee
Since we want to keep terms up to ${\cal O}(1/k)$ the holomorphic currents at points $z_2$ and $z_3$ must be contracted only
through the Abelian term of their OPE. The resulting correlators will involve the two primary fields and the three anti-holomorphic currents.
Next we employ the Ward identity for the current at the point $\bar z_1$. This current can not be contracted with the other anti-holomorphic currents through a {$\delta$-Kronecker term} because in such a case this term will be proportional either to $f_{abc} \delta_{ab}=0$ or to $f_{abc} \delta_{ac}=0$.
Also $\bar J_a(\bar z_1)$ can not be contracted with the external primary fields because in such a case the 
corresponding diagram will disconnected, thus it will be the product of
a bubble involving the points $z_2$ and $z_3$  times the rest of the diagram. Consequently, the only contribution that remains comes from the non-abelian contraction of either  $\bar J_a(\bar z_1)$ with either $\bar J_b(\bar z_2)$ or $\bar J_c(\bar z_3)$. In both cases the resulting diagrams will be disconnected, i.e. they will be the product of the tree-level $\langle \Phi \Phi\rangle$ correlator times a bubble involving all interactions points $z_i, \,\,\,i=1,2,3$.
Therefore, we conclude that
\be
J^{(2)}_{ii',jj'} =0\ .
\ee

\no
$\bullet$ The third term originates from the contraction of the current $J_a(z_1)$ with $J_b(z_2)$ through the Abelian term of their OPE
\be
J^{(3)}_{ii',jj'} = \int  {\text{d}^2z_{123}\ov z_{12}^2} \langle \Phi^{(1)}_{i,i'}(x_1,\bar x_1)
\bar J_a(\bar z_1)  \bar J_a(\bar z_2) J_c(z_3) \bar J_c(\bar z_3) \Phi^{(2)}_{j,j'}(x_2,\bar x_2)\rangle \ .
\ee
The last holomorphic current, i.e. the one at point $z_3$ should necessarily be contracted
with each of the external primaries giving a factor of $1/\sqrt{k}$ and leaving us with
a sum of two correlators
involving two primaries and the three anti-holomorphic currents.
Choosing the current at $z_1$ to be the one for which we will apply the Ward identity we obtain the following terms: \\
i) the term when $\bar J_a(\bar z_1)$ is contracted with $\bar J_a(\bar z_2)$. This diagram will have a factor of ${1 \ov  z_{12}^2 \bar z_{12}^2}$
indicating that it is disconnected and should, thus be ignored.\\
ii)  the term arising from the contraction of $\bar J_a(\bar z_1)$ with $\bar J_c(\bar z_3)$ through the non-Abelian term of their OPE will
also be zero since we have saturated the powers of $1/k$ and all remaining contractions should be Abelian resulting to the factor of $f_{acd} \delta_{ad}=0$.\\
iii) the term arising form the contraction of  $\bar J_a(\bar z_1)$ with $\bar J_c(\bar z_3)$ through the Abelian term of their OPE contributes that
\be
\label{more1}
\begin{split}
&-{1 \ov k}{1 \ov x_{12}^{2 \D_R} \bar x_{12}^{2 \bar \D_{R'}}} \int {\text{d}^2z_{123} \ov z_{12}^2 \bar z_{13}^2}
\Big({\big( t_a^{(1)} \otimes \tilde t_a^{(1) T} \big)_{ii',jj'}  \ov  (z_3-x_1)(\bar z_2-\bar x_1)}
+{\big( t_a^{(1)} \otimes \tilde t_a^{(2) } \big)_{ii',jj'} \ov  (z_3-x_1)(\bar z_2-\bar x_2)}+ \\
&+{\big( t_a^{(2) T} \otimes \tilde t_a^{(1) T} \big)_{ii',jj'}  \ov  (z_3-x_2)(\bar z_2-\bar x_1)}+
 {\big( t_a^{(2) T} \otimes \tilde t_a^{(2) } \big)_{ii',jj'}  \ov  (z_3-x_2)(\bar z_2-\bar x_2)}\Big)=
 2 {\pi^3\ov  k} {(t_a \otimes t^*_a)_{ii',jj'}\ov x_{12}^{2 \D_R} \bar x_{12}^{2 \bar \D_{R'}}}
\ln {\varepsilon^2\ov |x_{12}|^2} \, .
\end{split}
\ee
Notice that as always we keep the order of integrations. Furthermore, the first integral over $z_1$ gives a $\d^{(2)}(z_2-z_3)$ which make the
second integration over $z_2$ trivial. The third integral over $z_3$ is one of our basic ubiquitous ones.
\\
iv) the last contribution arises when $\bar J_a(\bar z_1)$ is contracted with the external primaries.
The corresponding integrals are
\be
\begin{split}
\label{more2}
&-{1 \ov k}{1 \ov x_{12}^{2 \D_R} \bar x_{12}^{2 \bar \D_{R'}}} \int {\text{d}^2z_{123} \ov z_{12}^2 \bar z_{23}^2}
\Big({\big( t_a^{(1)} \otimes \tilde t_a^{(1) T} \big)_{ii',jj'}  \ov  (z_3-x_1)(\bar z_1-\bar x_1)}
+{\big( t_a^{(1)} \otimes \tilde t_a^{(2) } \big)_{ii',jj'}  \ov  (z_3-x_1)(\bar z_1-\bar x_2)}+ \\
&+{\big( t_a^{(2) T} \otimes \tilde t_a^{(1) T} \big)_{ii',jj'}  \ov  (z_3-x_2)(\bar z_1-\bar x_1)}+
 {\big( t_a^{(2) T} \otimes \tilde t_a^{(2) } \big)_{ii',jj'}  \ov  (z_3-x_2)(\bar z_1-\bar x_2)}\Big)=
 2 {\pi^3\ov  k} {(t_a \otimes t^*_a)_{ii',jj'}\ov x_{12}^{2 \D_R} \bar x_{12}^{2 \bar \D_{R'}}}
\ln {\varepsilon^2\ov |x_{12}|^2} \, .
\end{split}
\ee
Adding the contributions from \eqref{more1} and \eqref{more2} we get for $J^{(3)}_{ii',jj'}$ that
\ba
J^{(3)}_{ii',jj'} =4 {\pi^3\ov  k} {(t_a \otimes t^*_a)_{ii',jj'}\ov x_{12}^{2 \D_R} \bar x_{12}^{2 \bar \D_{R'}}}
\ln {\varepsilon^2\ov |x_{12}|^2}\ .
\ea

\no
$\bullet$ The fourth term originates from the contraction of the current $J_a(z_1)$ with $J_c(z_3)$ through the non-Abelian term of their OPE.
It reads
\begin{equation}
J^{(4)}_{ii',jj'} = {f_{acd}\ov \sqrt{k}}\int {\text{d}^2z_{123} } \langle \Phi^{(1)}_{i,i'}(x_1,\bar x_1)
\bar J_a(\bar z_1) J_b(z_2) \bar J_b(\bar z_2) J_d(z_3) \bar J_c(\bar z_3) \Phi^{(2)}_{j,j'}(x_2,\bar x_2)\rangle
\end{equation}
Following the same steps as in the second contribution above one can show that
\begin{equation}
J^{(4)}_{ii',jj'} =0\ .
\end{equation}
\no
$\bullet$ The fifth term originates from the contraction of the current $J_a(z_1)$ with $J_c(z_3)$ through the Abelian term of their OPE
\begin{equation}
J^{(5)}_{ii',jj'} = \int {\text{d}^2z_{123} \ov z_{13}^2} \langle \Phi^{(1)}_{i,i'}(x_1,\bar x_1)
\bar J_a(\bar z_1) J_b(z_2) \bar J_b(\bar z_2)  \bar J_a(\bar z_3) \Phi^{(2)}_{j,j'}(x_2,\bar x_2)\rangle\ .
\end{equation}
Working as in the case of the third contribution we get that
\begin{equation}
J^{(5)}_{ii',jj'} =4 {\pi^3\ov  k} {(t_a \otimes t^*_a)_{ii',jj'}\ov x_{12}^{2 \D_R} \bar x_{12}^{2 \bar \D_{R'}}}
\ln {\varepsilon^2\ov |x_{12}|^2}\ .
\end{equation}
\no
$\bullet$ Finally, the last term originates from the contraction of the current $J_a(z_1)$ with the primary field $\Phi^{(2)}$
\begin{equation}
J^{(6)}_{ii',jj'} = -{1\ov \sqrt{k}}\int \text{d}^2z_{123} {(t^{(2)}_a)_j{}^k\ov z-x_2} \langle \Phi^{(1)}_{i,i'}(x_1,\bar x_1)
\bar J_a(\bar z_1) J_b(z_2) \bar J_b(\bar z_2) J_c(z_3) \bar J_c(\bar z_3) \Phi^{(2)}_{k,j'}(x_2,\bar x_2)\rangle\,.
\end{equation}
Following the same steps as in the first contribution one can show that
\begin{equation}
J^{(6)}_{ii',jj'} =2 {\pi^3\ov  k} {(t_a \otimes t^*_a)_{ii',jj'}\ov x_{12}^{2 \D_R} \bar x_{12}^{2 \bar \D_{R'}}}
\ln {\varepsilon^2\ov |x_{12}|^2}\ .
\end{equation}
Summing up all  six terms one obtains the final result at three-loops. It reads
\begin{equation}
\langle \Phi^{(1)}_{i,i'}(x_1,\bar x_1) \Phi^{(2)}_{j,j'}(x_2, \bar x_2)\rangle_\l^{(3)}
= -2 {\l^3\ov  k} {(t_a \otimes t^*_a)_{ii',jj'}\ov x_{12}^{2 \D_R} \bar x_{12}^{2 \bar \D_{R'}}}
\ln {\varepsilon^2\ov |x_{12}|^2}\ .
\end{equation}

\section{Perturbative computation of the $\langle J \Phi \Phi\rangle$ correlator}
\label{jff}

Finally, in this last appendix, we present the perturbative calculation of the  $\langle J \Phi \Phi\rangle$ three-point correlator.\\

{\bf Order ${\cal O}(\l)$:} This contribution is equal to
\begin{equation}
\begin{split}
\langle \Phi J\Phi\rangle^{(1)}&=-\frac{\lambda}{\pi}\int\text{d}^2z\,
\langle \Phi_{i,i'}^{(1)}(x_1)J_a(x_3)J_{a_1}(z)\bar J_{a_1}(\bar z)\Phi^{(2)}_{j,j'}(x_2)\rangle\\
&=-\frac{\lambda}{\pi}\int\text{d}^2z\,\frac{\langle \Phi_{i,i'}^{(1)}(x_1)\bar J_{a}(\bar z)\Phi^{(2)}_{j,j'}(x_2)\rangle}{(x_3-z)^2}\\\
&=-\frac{\lambda\,(\mathbb{I}_{R'}\otimes\tilde t^*_a)_{ii',jj'}}
{\sqrt{k}\,x_{12}^{2\Delta_R}\bar x_{12}^{2\bar\Delta_{R'}}}\,\left(\frac{1}{ x_{13}}-\frac{1}{ x_{23}}\right)\,,
\end{split}
\end{equation}
where we have used \eqref{JJf2}.

\no
{\bf Order ${\cal O}(\l^2)$:} This contribution is equal to
\ba
\label{2-loopjff}
&&\langle \Phi J\Phi\rangle^{(2)}=\frac{\lambda^2}{2!\pi^2}\int \text{d}^2z_{12}\,
\langle \Phi_{i,i'}^{(1)}(x_1)J_a(x_3)J_{a_1}(z_1)J_{a_2}(z_2)\bar J_{a_1}(\bar z_1)\bar J_{a_2}(\bar z_2)\Phi^{(2)}_{j,j'}(x_2)\rangle\nonumber\\
&&=\frac{\lambda^2}{2\pi^2}\int \text{d}^2z_{12}\,
\Big(\frac{\langle \Phi_{i,i'}^{(1)}(x_1) J_{a}( z_1)\Phi^{(2)}_{j,j'}(x_2)\rangle}{(x_3-z_2)^2\bar z_{12}^2}+
\frac{\langle \Phi_{i,i'}^{(1)}(x_1) J_{a}( z_2)\Phi^{(2)}_{j,j'}(x_2)\rangle}{(x_3-z_1)^2\bar z_{12}^2} \Big)\nonumber\\
&&=\frac{\lambda^2(t_a\otimes\mathbb{I}_R)_{ii',jj'}}
{2\sqrt{k}\,x_{12}^{2\Delta_R}\bar x_{12}^{2\bar\Delta_{R'}}}\,\left(\frac{1}{x_{13}}-\frac{1}{x_{23}}\right)\,,
\ea
where we have used \eqref{JJf1} disregarding bubble diagrams. Notice that
the second term in the second line of \eqref{2-loopjff} vanish,  since the
$z_1$ integration will give a $\d^{(2)}(x_3-z_2)$ which is set to zero in our regularization scheme.
Furthermore, notice that the order of integration is important. Had we changed this order the result of the
vanishing term would have been non-zero doubling the contribution of the first term in the second line of \eqref{2-loopjff}.

\no
{\bf Order ${\cal O}(\l^3)$:} This contribution is equal to
\begin{equation}
\begin{split}
&\langle \Phi J\Phi\rangle^{(3)}=\\
&-\frac{\lambda^3}{3!\pi^3}\int \text{d}^2z_{123}\,
\langle \Phi_{i,i'}^{(1)}(x_1)J_a(x_3)J_{a_1}(z_1)J_{a_2}(z_2)J_{a_2}(z_3)\bar J_{a_1}(\bar z_1)\bar J_{a_2}(\bar z_2)\bar J_{a_3}(\bar z_3)\Phi^{(2)}_{j,j'}(x_2)\rangle\\
&=-\frac{\lambda^3}{2\pi^3}\int \text{d}^2z_{123}\,
\frac{\langle \Phi_{i,i'}^{(1)}(x_1)\bar J_{a}(\bar z_3)\Phi^{(2)}_{j,j'}(x_2)\rangle}{(x_3-z_3)^2z_{12}^2\bar z_{13}^2}
=-\frac{\lambda^3\,(\mathbb{I}_{R'}\otimes\tilde t^*_a)_{ii',jj'}}
{2\sqrt{k}\,x_{12}^{2\Delta_R}\bar x_{12}^{2\bar\Delta_{R'}}}\,\left(\frac{1}{ x_{13}}-\frac{1}{ x_{23}}\right)\,,
\end{split}
\end{equation}
where we have used \eqref{JJf2}.

\end{appendices}

\end{document}